\documentstyle[12pt]{article}

\date{\today}

\def\be{\begin{equation}}
\def\ee{\end{equation}}
\def\bear{\begin{eqnarray}}
\def\eear{\end{eqnarray}}
\def\nn{\nonumber}

\def\wdg{{\wedge}}                              

\def\Re{{\rm Re\hskip0.1em}}
\def\Im{{\rm Im\hskip0.1em}}

\newcommand\px[1]{{\partial_{#1}}}

\newcommand\ppx[1]{{{\partial\over{\partial {#1}}}}}
\newcommand\pypx[2]{{{{\partial {#1}}\over{\partial {#2}}}}}
\newcommand\ppypxpx[3]{{{{\partial^2 {#1}}
                           \over{\partial {#2}\partial {#3}}}}}

\newcommand\tr[1]{{\mbox{tr}\{{#1}\}}}          
\newcommand\com[2]{{\lbrack {#1},{#2}\rbrack}}  

\def\BZ{{{\bf Z}}}

\def\BR{{{\bf R}}}
\newcommand\MR[1]{{{\bf R}^{#1}}}               
\newcommand\MC[1]{{{\bf C}^{#1}}}               
\newcommand\MS[1]{{{\bf S}^{#1}}}               
\newcommand\MT[1]{{{\bf T}^{#1}}}               
\newcommand\CP[1]{{{\bf CP}^{#1}}}              

\newcommand\SUSY[1]{{{\cal N}= {#1}}}           

\def\a{{\alpha}}
\def\b{{\beta}}
\def\g{{\gamma}}

\def\u{{\mu}}
\def\v{{\nu}}

\def\lam{{\lambda}}

\def\barz{{\overline{z}}}

\def\ZM{{\Xi}}   
\def\MTW{{\widetilde{{\cal M}}}}   
\def\Abl{{\cal A}} 
\def\Nil{{\cal N}} 
\def\Kom{{\cal K}} 

\def\RtSp{{\cal H}}    
\def\RtLat{{\Delta}}   
\def\PRtLat{{\Delta_{+}}}   

\newcommand\lrp[2]{{{\langle {#1}, {#2} \rangle}}} 

\def\bz{{\overline{z}}}
\def\bw{{\overline{w}}}
\def\sig{{\sigma}}
\def\blam{{\overline{\lambda}}}
\def\btau{{\overline{\tau}}}
\def\bsig{{\overline{\sigma}}}
\def\wlam{{\widetilde{\lambda}}}
\def\wtau{{\widetilde{\tau}}}
\def\wsig{{\widetilde{\sigma}}}
\def\bwlam{{\overline{\widetilde{\lambda}}}}
\def\bwtau{{\overline{\widetilde{\tau}}}}
\def\bwsig{{\overline{\widetilde{\sigma}}}}

\def\he{{\hat{e}}}

\begin{document}

\begin{titlepage}
\titlepage
\rightline{hep-th/9910236, PUPT-1895}
\rightline{\today}
\vskip 1cm
\centerline{{\Huge Effects in Gauge Theories}}
\centerline{{\Huge and a Harmonic Function on $E_{10}$}}
\vskip 1cm
\centerline{
Ori J. Ganor\footnote{origa@viper.princeton.edu}
}
\vskip 0.5cm

\begin{center}
\em  Department of Physics, Jadwin Hall \\
Princeton University\\
NJ 08544, USA
\end{center}

\vskip 0.5cm

\abstract{
In a previous paper we conjectured that
the structure of various gauge theories as well as M-theory on $\MT{8}$
is encoded in a unique
function $\Xi$ on the coset $E_{10}(\BZ)\backslash E_{10}(\BR)/K$
and that this function is harmonic with respect
to the $E_{10}(\BR)$ invariant metric.
In this paper we elaborate on the conjecture.
We discuss various mass deformations of the D-instanton integral
and their realizations in $\Xi$. We then present a conjectured
prescription for extracting partition functions of the twisted
little-string theory out of $\Xi$.
We also study various effects of combinations of branes such as
D0-branes near D4-branes with 2-form flux, D-instantons near
Taub-NUT metrics, and more,
in terms of harmonic functions on $E_d(\BR)/K$.
We propose tests of the conjecture that are related to BPS states
of global symmetries in gauge theories.
}
\end{titlepage}
             

\tableofcontents

\section{Introduction}\label{intro}
In a previous paper \cite{GanXi}, we conjectured that there exists
a unique function $\ZM$ that encodes the partition functions
of a large class of gauge theories.
The function $\ZM$ was defined on the coset space 
$\MTW = E_{10}(\BZ)\backslash E_{10}(\BR)/\Kom$ of the group $E_{10}(\BR)$
(the exponentiation of the Lie algebra $E_{10}$) by the maximal
compact subgroup $\Kom$ on the left and a discrete subgroup $E_{10}(\BZ)$
on the right.
We proposed that various field-theoretic partition functions can
be extracted from $\ZM$ by appropriate Fouri\'er transforms
with respect to periodic variables of $\MTW$. Furthermore, 
we proposed that a partition function of M-theory on $\MT{9}$ might
be well-defined if we include generic transverse $SO(2)$-twists
and that this partition function is also encoded in $\ZM$.
(See \cite{GanXi} for references on previous works related to $E_{10}$.)

The purpose of the present paper is to refine the conjecture.
The basic idea is the connection between Euclidean branes that wrap
cycles in M-theory on $\MT{d}$ and positive roots of $E_d$ ($d=1\dots 8$).
In $\MTW$, and of course also in the moduli space,
$E_{d(d)}(\BZ)\backslash E_{d(d)}(\BR)/\Kom_d$, of M-theory on $\MT{d}$,
a positive root is related to a periodic variable $\phi$.
An instanton made of $N$
wrapped Euclidean branes comes with a characteristic factor of 
$e^{i N\phi}$ \cite{BBS,WitNON}.
Thus, by extracting the terms in $\ZM$ that behave
as $e^{i N\phi}$ we can extract information about the gauge (or other)
theory associated with $N$ branes at low-energy.
This is, in general, a theory with 16 supersymmetries and noncompact moduli
whose partition function is not well-defined.
For example, the D-instanton action is an integral with
$10N$ non-compact modes.
To get a well-defined function we have to augment the partition
function with mass-terms that break the supersymmetry and get rid of
the moduli.
For example, in the case of the D-instanton action, the mass-terms are
quadratic in the variables.
 How do we interpret this augmentation in terms of variables
of $\MTW$?

We will argue that the general procedure is as follows.
We have to find another periodic variable, $\psi$ in $\MTW$, that
will be mapped to the coefficient of the term in the deformed
action (in the D-instanton example this would be proportional to the
mass). However, as we will see, setting $\psi$ to the desired value
is not enough, by itself.
We have to find a third periodic variable, $\chi$, that will ``connect''
$\psi$ to the action. The prescription will then be to extract
out of $\ZM$ the term that behaves as $e^{iN\phi + i\chi}$ and study
it as a function of $\psi$.
More generically, we have to identify pairs of variables, $\psi_j$
and $\chi_j$, ($j=1,\dots$) and isolate out of $\ZM$ the term
that behaves as $e^{iN\phi + i\sum\chi_j}$. We then have to study
it as a function of the deformations $(\psi_1,\psi_2,\dots)$.
As will be reviewed below,
the variables $\psi_j$ and $\chi_j$ correspond to positive roots, $\b_j$
and $\g_j$,  of $E_{10}$. We will call $\b_j$ the {\em ``hook''}
and $\g_j$ will be its corresponding {\em ``bait''}.

The motivation for this procedure is that deformations of
the theories with 16 supersymmetries that describe the dynamics of
the branes \cite{WitBND,StrOPN,SeiVBR} can be realized by inserting
other objects near the branes \cite{BDS,SeiIRD}.
The most important case will be when the ``bait'' $\g$ is orthogonal,
in the root lattice of $E_{10}$, to the root $\a$ which corresponds to
$\phi$ and the ``hook'' $\b$
is at $60^\circ$ to $\a$ and $120^\circ$
to $\g$.
 (In other words, $\lrp{\a}{\g}=0$ and $\lrp{\a}{\b}=-\lrp{\g}{\b}=1$
where $\lrp{\cdot}{\cdot}$ is the inner product in the weight lattice
of $E_{10}$.)
 This system corresponds to a deformation that preserves
half the supersymmetry and has various manifestations.
Among them are the compactifications with R-symmetry twists \cite{CGKM},
the elliptic models of \cite{WitBR}, a D-instanton inside a D3-brane
\cite{Douglas}
with background NS-NS 2-form flux and more.
Another interesting case is when $\lrp{\a}{\g}=-2$.
 This system corresponds to D0-branes near D8-branes
\cite{BSSil,DFK,BGL}, and some other cases as well.

In \cite{GanXi} we suggested the above procedure for extracting
partition functions of gauge theories out of $\Xi$ but we did not
specify exactly which {\em hooks} and {\em baits} need to be chosen
for a particular theory.
One of the goals of the present paper is to identify them
more precisely. We use as a case-study the D-instanton action
that is an integral over $10$ $N\times N$ matrices.
We will suggest a set of hooks and baits that correspond to mass
deformations of the D-instanton action that break supersymmetry
completely and lift all the flat directions.
The result of the integration should therefore be a nontrivial
function of the deformation parameters and we propose that it is
encoded in a Fouri\'er transform of $\ZM$.

In this paper we will adhere to the interpretation of \cite{GanXi}
and consider only the deformed theories with well-defined
partition functions (i.e. with all the flat directions lifted by
mass deformations).
However, our discussion is also relevant to the study of
instanton effects in M-theory on $\MT{8}$ and lower dimensional
tori (see \cite{OPRev,OPRevSh,OPNew} and refs. therein).
The $R^4$ term, which is also related 
to 16-fermion, $\lambda^{16}$, terms by supersymmetry
\cite{GGI,PiolH,GreSet},
can be calculated from single-weight instantons
(i.e. terms made from a single BPS-brane wrapped $N$ times).
The instantons corresponding to Euclidean
${1\over 4}$-BPS states give rise to
terms of the form $H^{4g-4} R^4$ \cite{BerVaf} where $H$
is an RR field strength. They are likely to
be related by supersymmetry to terms of
the form $e^{i N\phi + i k\chi}\lambda^{24}$ where $\lambda^{24}$
is a shorthand for a $24$-fermion term and $\phi$ and $\chi$
are two periodic phases in the moduli space.
They correspond to roots $\a$ and $\g$ that satisfy $\lrp{\a}{\g}=0$.
The power of $H$ in $H^{4g-4} R^4$ is related to $k$ and $N$ via
$g=N k$. Similarly, instanton configurations that 
preserve $2^r$ supersymmetries
are likely to contribute to terms of the form
 $e^{i N\phi+i\sum_1^r k_j\chi_j}\lam^{32-2^r}$.

The paper is organized as follows.
In section (\ref{roots}) we review the relation between instantons and
positive roots of $E_{d}$ and between harmonic functions on
$E_{d(d)}/\Kom$ and the action of the instantons.
(See also the comprehensive reviews in \cite{OPRev,OPRevSh,OPNew}.)
In section (\ref{hooks}) we discuss the relation between mass-like
terms in the instanton actions and hooks and baits in $\ZM$.
We discuss various U-dual systems that demonstrate this principle.
We also conjecture that the generic action of a BPS instanton
in M-theory on $\MT{8}$ (preserving ${1\over {16}}$ of the supersymmetry)
is a harmonic function on $E_{8(8)}/SO(16)$.
In section (\ref{twhook}) we restrict to the case of the D-instanton
integral and we proceed to study mass deformations that preserve
${1\over 4}$ or less of the supersymmetry.
They are realized by two or more hooks (and their corresponding baits).
In section (\ref{mintwo}) we return to systems made up of a pair of
BPS instantons that correspond to roots $\a$ and $\b$ with inner
product $\lrp{\a}{\b}=-2$.
 Whereas the pairs corresponding to roots $\a$ and $\b$
with $\lrp{\a}{\b}=0$ have been extensively studied the pairs
with $\lrp{\a}{\b}=-2$, that also preserve half the supersymmetry,
have been studied less. We briefly discuss a particular case
of a D(-1)-brane near a D7-brane and suggest that other U-dual systems
might be interesting to study.
In section (\ref{bmintw}) we study the effect of baits with
$\lrp{\a}{\b}=-2$ and its relation to supersymmetry breaking.
In section (\ref{higherd}) we suggest various models for 
extracting the partition function of higher dimensional theories.
We briefly discuss an example where some of the BPS particle spectrum
can be manifested.
In section (\ref{groups}) we study the Laplacian on $E_{10}$
in conjunction with the conjecture that $\ZM$ is harmonic
and given the proposed procedure for mass deformations.


\section{Instantons, Branes, and Positive Roots}\label{roots}
The moduli space of M-theory on $\MT{d}$ is given by
$E_{d(d)}(\BZ)\backslash E_{d(d)}(\BR)/\Kom$ where $\Kom$ is a maximal
compact subgroup. An element $g\in E_{d(d)}(\BR)/\Kom$ can be decomposed
as $g = n\circ a$ where $a\in (\MR{+})^{d}$ is an element in
a maximal abelian subgroup and $n\in \Nil$ is an element in a
nilpotent subgroup $\Nil$.
For example, for $d=8$ we have $\Kom=SO(16)$ and if $\MT{8}$ is of
the form $(\MS{1})^{8}$ with no fluxes of the 3-form or dual 6-form
and no VEVs to the 2+1D duals of the vectors then $a$ can be taken
as the vector $(R_1,\dots,R_8)$ of the 8 radii of $\MS{1}$.
The elements of $n$ contain all the other moduli, i.e. fluxes, Dehn
twists and duals of vectors. These become periodic phases after
modding out by $\Nil\bigcap E_{8(8)}(\BZ)$.

\subsection{Single instantons}
Various terms in the low-energy effective action of M-theory
on $\MT{8}$ receive contributions from 2+1D space-time instantons.
The simplest of these instantons can be described by taking a BPS particle 
of M-theory on $\MT{7}$ with a Euclidean world-line along the remaining
cycle of $\MT{8}$.
These instanton terms have a characteristic coefficient of the form,
$e^{-2\pi T + 2\pi i\phi}$
where $T$ is the action of the instanton and $\phi$ is the phase that
couples to it.
Restricting to $\MT{8}$'s of the form $(\MS{1})^8$ we find 4 kinds of
BPS instantons:
\begin{itemize}
\item
KK states with Euclidean world lines with $T = R_i R_j^{-1}$ ($i\neq j$).
\item
Wrapped membranes with $T=\prod_{k=1}^3 R_{i_k}$.
\item
Wrapped fivebranes with $T=\prod_{k=1}^6 R_{i_k}$.
\item
KK monopoles with $T=R_{i_8}^2 \prod_{k=1}^7 R_{i_k}$.
\end{itemize}

For each of those instantons, the phase $\phi$ is one periodic variable
in $\Nil$ and hence corresponds to a positive root $\a$ in the root
lattice $\RtLat$ of $E_8$. The tension $T$ can then be calculated
as follows. Identify,
$$
\lam = (\log R_1,\dots,\log R_8)
$$
as a vector in the coroot space $\RtSp$ of $E_8$. Let $\lrp{\cdot}{\cdot}$
be the inner product (using the Cartan matrix).
The tension is then given by,
$$
T_\a = e^{\lrp{\a}{\lam}}.
$$
It is interesting to note that the factor $e^{-2\pi T_\a}$ can also
be determined by looking for a harmonic function on the moduli space
that behaves as $e^{2\pi i\phi_\a}$. Up to prefactors, the function
$e^{-2\pi T_\a + 2\pi i\phi_\a}$ is harmonic!

\subsection{Pairs of instantons}
Let $\a,\b\in\PRtLat$ be postitive roots (here $\PRtLat$
is the set of positive roots).
There are certain terms in the low-energy effective action
that receive contributions from BPS instantons and behave as
$e^{-2\pi T_\a + 2\pi i\phi_\a}$ and $e^{-2\pi T_\b + 2\pi i\phi_\b}$.
We will now discuss terms that behave as
$$
e^{-2\pi T + 2\pi i k\phi_\a + 2\pi i m\phi_\b},
$$
where $T$ is a real function of the moduli.
We will not be very specific about whether these are 16-fermion terms
or something else. More important for us will be the behavior of $T$.
Given $T_\a$ and $T_\b$, the behavior is determined by the product
$\lrp{\a}{\b}$ of the roots in the weight lattice.

Before we proceed let us present two formulas for calculating
$\lrp{\a}{\b}$.
Since each positive root corresponds to an instanton, it is convenient
to characterize $\a$ by the vector of integers $(n_1,\dots, n_d)$
such that the action of the instanton on $(\MS{1})^d$ is given by,
$$
T_\a = \prod_{i=1}^d R_i^{n_d}.
$$
If $\b$ is similarly characterized by $(m_1,\dots,m_d)$ then,
$$
\lrp{\a}{\b} =
  \sum_{i=1}^d n_i m_i 
   -{1\over 9}\left(\sum_i n_i\right)\left(\sum_j m_j\right).
$$
We will in general use $d=8$ but later on it will be necessary to
extend this to $d=10$ that formally corresponds to M-theory

Sometimes it will be more convenient to express the roots in the
type-II language.
Suppose we compactify type-IIA (or type-IIB) on $(\MS{1})^{d-1}$ with
radii of lengths $l_1,\dots,l_{d-1}$ in string units and
a string coupling constant $\lam$.
A positive root $\a$ can be characterized by the numbers 
$(p,s_1,\dots,s_{d-1})$ such that,
$$
T_\a = \lam^{-p}\prod_{i=1}^{d-1} l_i^{s_i}.
$$
If we take another root $\a'$ with,
$$
T_{\a'} = \lam^{-p'}\prod_{i=1}^{d-1} l_i^{s_i'},
$$
then product is then given by,
$$
\lrp{\a}{\a'} = 2p p' -{1\over 2}p\sum s_i' -{1\over 2}p'\sum s_i
    +\sum s_i s_i'.
$$
Note that T-duality on the $k^{th}$ direction acts as:
$$
s_k\rightarrow p-s_k,
$$
leaving $p$ and all the other $s_i$'s intact.
S-duality of type-IIB, on the other hand, keeps all the $s_i$'s intact
but changes:
$$
p\rightarrow {1\over 2}\sum s_i -p.
$$
It is amusing to note that in higher dimensions there are exotic
``branes'' that correspond to roots $|\a|^2=2$ (see \cite{EGKR}).
Some of them are invariant under S-duality.
For example, in addition to the D3-brane and KK-monopole
that are invariant, we have
the formal object with action
${1\over {\lambda^3}}l_1 l_2 l_3 l_4 l_5 l_6 l_7^2 l_8^2 l_9^2$.

Before we discuss the various combinations of two instantons let us
mention one more mathematical detail.
Suppose $\a,\b\in\Delta_{+}$ are such that $\a+\b\in\Delta_{+}$
is also a root.
The three periodic variables $e^{2\pi i\phi_\a}$, $e^{2\pi i\phi_\b}$ and
$e^{2\pi i\phi_{\a+\b}}$ do not parameterize $\MT{3}$ but rather 
$e^{2\pi i\phi_{\a+\b}}$ is a section of an $\MS{1}$ bundle of
first Chern class $c_1 = 1$ over the $\MT{2}$ parameterized
by $e^{2\pi i\phi_\a}$ and $e^{2\pi i\phi_\b}$
(see \cite{GanXi} and refs therein).

Given an instanton contribution of the form
\be\label{tphiab}
e^{-2\pi T + 2\pi i N\phi_\a + 2\pi i K\phi_\b},
\ee
we would like to ask how $T$ behaves as a function of $T_\a$ and $T_\b$.
To be rigorous, we have to be more specific about the other phases.
In general, if $\a-\b$ (or $\b-\a$) is a positive root we have
to set $\phi_{\a-\b} = 0$ because $e^{2\pi i\phi_\a}$ is a section
of a nontrivial line bundle, as we explained above.
As for the other $\phi_\gamma$'s, we can assume that the expression
is independent of them.
There are various cases according to the value of $\lrp{\a}{\b}$.

\begin{itemize}
\item
If $\lrp{\a}{\b} = 1$ then $T=\sqrt{N^2 T_\a^2 + K^2 T_\b^2}$
and (\ref{tphiab}) is the contribution of a BPS instanton.
For example, $\a$ might correspond to a Euclidean D1-brane wrapped on the
$1^{st}$ and $2^{nd}$ directions and $\b$ might correspond to
a D1-brane wrapped on the $1^{st}$ and $3^{rd}$ directions.
Then there exists a single BPS D1-brane wrapped on the $1^{st}$ direction
and the diagonal of the torus made from the $2^{nd}$ and $3^{rd}$ 
directions.
For another example, $\a$ might correspond to a D(-1)-brane
and $\b$ might correspond to a D1-brane which combine to a D1-brane
with electric flux (T-dual to a D0-brane and a D2-brane).
It is interesting to note that the functional behavior 
$T=\sqrt{N^2 T_\a^2 + K^2 T_\b^2}$ also comes out from the
leading order behavior
of a harmonic function on the moduli space that behaves as
(\ref{tphiab}).

\item
If $\lrp{\a}{\b} = 0$ then $T=N T_\a + K T_\b$ and (\ref{tphiab}) is
the contribution of a ${1\over 4}$BPS instanton.
For example, $\a$ might correspond to a D(-1)-brane (with
action ${1\over \lam}$) and $\b$ to a 
D3-brane (with action ${1\over \lam}l_1\cdots l_4$).
For another example, $\a$ might correspond to an M5-brane
(with action $R_1\cdots R_6$)
and $\b$ to a KK-monopole (with action
$R_1\cdots R_7 R_8^2$) that engulfs the M5-brane.
A third example is a D4-brane (with action ${1\over \lam}l_1\cdots l_5$)
and an NS5-brane (with action ${1\over {\lam^2}}l_1\cdots l_4 l_6 l_7$)
that intersect along a 4-dimensional hyper-plane.
A fourth, U-dual example is furnished by $N$ M2-branes
(with action $R_1 R_2 R_3$) intersecting $K$ M2-branes
(with action $R_1 R_4 R_5$).
 Once again, the behavior
$T=N T_\a + K T_\b$ also comes out from the leading order behavior
of a harmonic function on the moduli space that behaves as
(\ref{tphiab}).
These combinations of instantons contribute to terms
of the form $H^{4g-4} R^4$ in the low-energy description 
of M-theory on $\MT{d}$, where $H$ is an appropriate field-strength
of a low-energy field and $g = k N$ \cite{BerVaf}.

\item
If $\lrp{\a}{\b} = -1$ then the instanton again preserves ${1\over 2}$
of the supersymmetry and has action $\sqrt{N^2 T_\a^2 + K^2 T_\b^2}$.
One example is furnished by a D(-1)-brane and a D5-brane.
This is T-dual to the system of a D0-brane and a D6-brane studied
in \cite{WatZS}.
Supersymmetry is broken when the D0-brane and D6-brane
are far from  each other. However, type-IIA on $\MT{6}$
actually has a BPS particle that has the same charge of a D0-brane
and a D6-brane. To see this, recall that type-IIA on $\MT{6}$
has an $SL(2,\BZ)$ duality group (a subgroup of the full $E_7(\BZ)$
U-duality) that acts on $\tau = {{i V}\over {\lam^2}} + \chi$.
Here $V$ is the volume of $\MT{6}$ (in string units) and $\lam$ is
the 10D string coupling constant. The periodic modulus $\chi$ is the axion
(dual to the NSNS 2-form). The
S-duality $\tau\rightarrow -1/\tau$ transforms a D0-brane
into its dual, the wrapped D6-brane.
The transformation $\tau\rightarrow\tau+1$ transforms a D6-brane
into an object with the charges of a D0-brane and a D6-brane together.
Another way of obtaining this ``dyonic''
object is by starting with a wrapped D6-brane
and quantizing the collective coordinate corresponding to
rotations of the $11^{th}$ (M-theory) direction.
There is no contradiction between these statements and the results of
\cite{WatZS} because there the systems had more charges.
The behavior $T=\sqrt{N^2 T_\a^2 + K^2 T_\b^2}$ can also be deduced
from a harmonic function, since the Laplacian is defined to be U-duality
invariant. However, as a part of a harmonic function $T$ would also
depend on $\phi_{\a+\b}$ and the expression
$\sqrt{N^2 T_\a^2 + K^2 T_\b^2}$ is obtained only when we set 
$\phi_{\a+\b}$ to zero.

\item
If $\lrp{\a}{\b} = -2$ the instanton is again ${1\over 4}$BPS and
$T=N T_\a+K T_\b$.
Unlike the previous BPS cases, it is not obvious
that this relation does not  seem 
to be directly related to a harmonic function.
However, this case is more complicated for various reasons.
First note that $|\a+\b|^2 = 0$. This means that the Cartan matrix
is either semidefinite or indefinite. Thus, we must have $d\ge 9$ which
means that we are dealing with (the abstract)
compactification to 1+1D or less.
The characterization of the root lattice of $E_9$ and $E_{10}$
(see \cite{Kac}) then implies that $\a+\b$ is also a positive root.
These systems will be discussed section (\ref{mintwo}).

\end{itemize}

\subsection{Harmonic functions and BPS actions}
As we have seen in the examples, the exponentials of the
actions of ${1\over 2}$-BPS
and ${1\over 4}$-BPS instantons are harmonic functions on the 
moduli space.
The most generic statement of this sort would be that the exponential
of the action
of a generic BPS instanton on $\MT{8}$ (that preserves only
${1\over {16}}$ of the supersymmetry) is given by a harmonic
function on $E_{8(8)}(\BZ)\backslash E_{8(8)}(\BR)/SO(16)$.
I will not attempt to prove this statement here.
However, let us outline a possible direction.
The idea is to find 60 complex linear differential operators
${\cal L}_i$ that annihilate the action of an instanton and
such that the Laplacian on the moduli space can be written as
$\nabla = \sum {\cal L}_i^\dagger {\cal L}_i$.
For each set of instanton charges the set of ${\cal L}_i$'s
could be different.
These operators are generalizations of the statement that, for
example, the instanton action of 4D Yang-Mills theory is holomorphic
in $\tau = {{8\pi i}\over {g^2}} + {\theta\over {2\pi}}$.
Given the two supersymmetry generators that preserve the instanton
charges, we can construct 1-forms of the form $A_i(\phi) d\phi^i$
(with $i=1\dots 120$) on the moduli space
that should remain constant in an instanton configuration
(i.e. $A_i(\phi)\px{\u}\phi^i = 0$).
The operators ${\cal L}_i$ can be constructed as a basis for
the orthogonal space to these 1-forms.

If the conjecture that the generic BPS instanton of M-theory
on $\MT{8}$ is described by a harmonic action is true then
a lot of information about BPS states in gauge theories can be 
extracted from it.
For example, in section (\ref{higherd}) we will construct 
instantons that are described by mass deformed $\SUSY{4}$ SYM.
We will argue that the mass of the adjoint scalar can be extracted
from the instanton action by considering an extra charge.
This resulting instanton can be embedded in M-theory
on $\MT{8}$ and, according to the conjecture, could be described
by a harmonic function.


\section{``Catching'' deformations of gauge theories}\label{hooks}
Let $\a_0$ be the root corresponding to the D(-1)-brane.
The ``action'' is $T_{\a_0} = {1\over {\lam}}$ and the phase
is $\phi_{\a_0} = \chi$, the RR partner of the dilaton.
The contribution of $N$ D(-1)-branes to low-energy processes is of
the form,
$$
Z e^{-2\pi N T_{\a_0} + 2\pi i N \chi}.
$$
In appropriate asymptotic regions of the moduli space, the prefactor
$Z$ can be calculated from the action,
\be\label{Izero}
I_0 = -{1\over 4}\tr{\com{X_I}{X_J}\com{X^I}{X^J}} 
+\Gamma^I_{\a\b}\tr{\psi^\a\com{X_I}{\psi^\b}},
\qquad I,J=1\dots 10,\qquad \a,\b=1\dots 64.
\ee
Here $X$ are measured in Einstein units,
$\Gamma^I_{\a\b}$ are Dirac matrices of $SO(10)$,
$X_I$ are $N\times N$ hermitian matrices in
the vector representation of $SO(10)$ and $\psi^\a$ are
$N\times N$ hermitian matrices with anti-commuting elements in the
spinor representation of $SO(10)$.
To get a specific quantity one must insert
certain couplings to the background fields and
integrate over the $X$'s and $\psi$'s (as in \cite{GGI}).

Now consider the modified action,
\be\label{Imass}
I = I_0 + (M^2)_{IJ} X^I X^J + m_{\a\b}\psi^\a\psi^\b.
\ee
We would like to realize such deformations in terms of instantons
in M-theory on $\MT{d}$.

\subsection{Hooks and baits}
The general idea is to map a mass term $m$ to a phase $\phi_\b$
for an appropriate positive root $\b\in\PRtLat$ such that
$m = c\phi_\b$ in the limit $\phi_\b\rightarrow 0$ and $c\rightarrow\infty$
is a function of the radii $R_1\dots R_d$.
We will see that in order to execute the plan we need another root
$\g\in\PRtLat$ and then we have to consider terms that behave as,
$$
(\cdots) e^{2\pi i N \phi_{\a_0} + 2\pi i\phi_\g}.
$$
For appropriately chosen $\b,\g\in\PRtLat$, and in appropriate
asymptotic regions of the moduli space, the prefactor
$(\cdots)$ will be calculated from the massive D-instanton integral.
We will call the root $\g$ the {\em ``bait''} and the root $\b$ will
be called the {\em ``hook''}.
We will study several examples below.
The examples will be:
\begin{itemize}
\item
A D-brane inside a KK-monopole with a twist.
\item
The elliptic brane configurations of \cite{WitBR}.
\item
A D-instanton near a KK-monopole with a B-field turned on.
\item
A D-instanton near a D3-brane with a B-field turned on.
\end{itemize}

In these examples it will turn out that we need the roots to satisfy
\be\label{azbg}
\lrp{\a_0}{\g} = 0,\qquad
\lrp{\a_0}{\b} = 1,\qquad
\lrp{\b}{\g} = -1.
\ee
The low-energy description of each example is different.
However  the examples are U-dual to each other and the main point
is that once $\phi_\b\neq 0$, there is a ``bound'' configuration
with action $T_{\a_0} + T_\g$ whereas if the instantons are
separate, their action is 
\be\label{unbndac}
T_\g + T_{\a_0} \sqrt{1 + \left({{\phi_\b}\over {T_\b}}\right)^2}
> T_\g + T_{\a_0}.
\ee
This forces the instanton with action $T_{\a_0}$ to be
at the center of the instanton with action $T_\g$ and
creates an effective mass term for the separation mode.
(See also \cite{SenUD,SenMA}).

The formula $T_1\equiv T_{\a_0} + T_\g$ for the action of the bound state
is actually only valid in a certain region of the moduli space.
In another region, when the competing
\be\label{Ttwo}
T_2 \equiv T_{\a_0} \sqrt{1 + \left({{\phi_\b}\over {T_\b}}\right)^2}
  + {{T_\g}\over
  {\sqrt{1 + \left({{\phi_\b}\over {T_\b}}\right)^2}}}
\ee
becomes smaller, there is a ``phase-transition'' to 
the other action $T_2$.
This phenomenon is well-known for $(p,q)$-string networks
on a slanted $\MT{2}$ \cite{BerKol}.
In this case we can take $T_{\a_0} = {1\over {\lam}}l_2 l_1$
and $T_\g = R_3 R_1$. Here $R_2 R_3^{-1} = \Im\tau$
and we take $\phi_\b = \Re\tau$.
When $\lam$ is small, $T_1$ is the correct formula for
the action of the string network. When $\lam$ is large 
$T_2$ is the correct formula.

\subsection{The twisted instanton actions}\label{dzkk}
The first example is a modification of \cite{CGKM}.
Take a Euclidean D0-brane that corresponds to a root $\a_0$ with action
$\lam^{-1}l_1$. Now embed it inside a KK-monopole with action
$\lam^{-2}l_7^2 l_1\cdots l_6$. This will be the bait-root $\g$.
The hook-root $\b$ will be the Dehn twist of the circle in the $7^{th}$ 
direction as we go around the $1^{st}$ direction.
It corresponds to an action $l_1 l_7^{-1}$.
These roots satisfy (\ref{azbg}).
When the Dehn twist is small it acts as an effective mass term
to 4 out of the 9 zero modes of the fields of the D0-brane just
as in \cite{CGKM}.
$M^2$ in (\ref{Imass}) has eigenvalues proportional to:
$$
M^2 \sim 
4\left\{ \phi_\b^2 \right\},\,
6\left\{ 0 \right\}.
$$

Let us discuss in what regions of moduli space the approximation 
of a massive 0D integral is valid. Let $\zeta$ be the vector
$(\log R_1,\dots,\log R_d)$.
The D0-brane action is a good approximation when the
string coupling constant $\lam\rightarrow 0$.
Let $X$ be a generic variable in the D-instanton
integral, measured in string units, such that the D-instanton action
is proportional to ${1\over {\lam}}$.
If $X$ is measured in string units, the D0-brane action is, schematically,
${1\over {\lam}}\int(\dot{X}^2 + X^4)$.
Corrections in $\a'$ behave as ${1\over {\lam}} X^{k}$ where $k\ge 6$.
String loop corrections give even smaller contributions.
Let us first suppress the time dependence.
The order of magnitude of the zero mode $X_0$  is
 $X_0\sim \lam^{1/4}l_1^{-1/4}$.
The compactified time interval is $l_1$.
The Fourier modes of $X$ along $l_1$ have a quadratic 
term of the form ${{n^2}\over {\lam l_1}}X_n^2$ which implies
$X_n\sim \lam^{1/2} l_1^{1/2}$.
These fluctuations can be ignored if
$\lam^{1/2} l_1^{1/2}\ll \lam^{1/4}l_1^{-1/4}$
so we require $l_1\ll\lam^{-1/3}$.
For the KK-monopole, we want the fluctuations in $X$ to be small 
compared to $l_7$. Thus, $\lam^{1/4}l_1^{-1/4}\ll l_7$.

To summarize we have,
$$
\lam^{1/4}l_1^{-1/4}\ll l_7,\qquad
\lam\ll 1,\qquad l_1\ll \lam^{-1/3}.
$$
We also need to require $\lam^{1/4}l_1^{-1/4}\ll l_i$ for all $i\neq j$.

\subsection{Elliptic brane configuration}\label{ellip}
Another U-dual example is the elliptic model of \cite{WitBR}.
Take a Euclidean D0-brane with action ${1\over \lam}l_1$ (corresponding
to the root $\a_0$)
and an NS5-brane with action ${1\over {\lam^2}}l_2\cdots l_7$
(corresponding to $\g$).
Now add a Dehn twist that corresponds to the root
$l_2^{-1} l_1$ (corresponding to $\b$).
According to the arguments of \cite{WitBR}, at low-energies
the action induced on the D0-brane, dimensionally reduced to 0D,
is of the form (\ref{Imass}).
$M^2$ has eigenvalues proportional to:
$$
M^2 \sim
4\left\{ {{\phi_\b^2}\over {l_2^2}} \right\},\,
6\left\{ 0 \right\}.
$$

Note that in this example (\ref{unbndac}) is satisfied as follows.
If we separate the Euclidean D0-brane from the NS5-brane,
with the Dehn twist $\phi_\b$ turned on, the length of the world-line of
the D0-brane will be $\sqrt{l_1^2 + l_2^2\phi_\b^2}$ which is
bigger than $l_1$.

\subsection{D-instanton inside a D3-brane}
Let us recall the system of a D(-1)-brane near a D3-brane.
Let us first take the D3-brane to be the primary root $\a_0$
and the ``hook'' root will be an NSNS 2-form flux.
\be
T_{\a_0} = {1\over \lam} l_1 l_2 l_3 l_4,\qquad
T_\g     = {1\over \lam},\qquad
T_\b     = l_1 l_2.
\ee
The phase is $\phi_\b = B_{12} l_1 l_2$.
Now take $l_1,l_2,l_3,l_4\rightarrow 0$.
This yields the construction of \cite{CDS,DH} of Yang-Mills
theories on a noncommutative torus. The D3-brane becomes
an instanton \cite{Douglas} of the noncommutative theory.
Because of the $(F-B)^2$ term in the action of the D3-brane,
the action of the unbound system is bigger by:
$$
{1\over {2\lam}} B_{12}^2 l_1 l_2 l_3 l_4 = 
T_{\a_0}\left({{\phi_\b}\over {T_\b}}\right)^2,
$$
to lowest order in $B_{12}$.

On the other hand we can take the D(-1)-brane to be the primary
root, $\a_0$, and the D3-brane to be the hook $\g$.
In the limit of \cite{SWncg} this system would become
an instanton of $U(1)$ SYM with the noncommutativity set
by $B_{12}$.
In this case, we must use formula (\ref{Ttwo}).
The D(-1)-brane has an action of ${1\over {\lam}}$ outside
the D3-brane. Inside the D3-brane it has an effective coupling
constant of (see eqn (2.44) of \cite{SWncg}):
$$
\lam \left( {{\det (G+2\pi \a B)}\over {\det G}} \right)^{1/2}
$$
where $G$ is the metric.
Thus the action of the D(-1)-brane is smaller in the bound state,
in accord with (\ref{Ttwo}).

At weak coupling and for large $B$ field,
the system is described by an integral on the moduli
space of noncommutative instantons.
For small $B$ fields, the  system is described by a matrix-model
with fundamental hyper-multiplets and a 
Fayet-Illiopoulos term proportional to the $B$ field
(see \cite{SWncg} and references therein).

\subsection{A graviton trapped in a string}
Let us describe yet another example of the same kind.
This example is likely to contribute to $24$-fermion terms
in the low-energy effective action of type-II string theory on
$\MT{2}$ in 8-dimensions.
The actions are as follows:
\be
T_{\a_0} = l_1 l_3^{-1},\qquad
T_\g     = {1\over \lam} l_1 l_3,\qquad
T_\b     = l_2 l_3^{-1}.
\ee
This system describes the bound state of a string and graviton.
The string is stretched on one of the cycles of 
a slanted $\MT{2}$. If $\tau=\tau_1 + i\tau_2$ is the complex structure
of the $\MT{2}$ then $\phi_\b=\tau_1$.
The bound state describes a string carrying momentum in one direction.
The bound state has an energy gap because the graviton will have energy
proportional to $\tau_2^{-1}$ outside the string, but only
$|\tau|^{-1}$ inside the string.

\subsection{An instanton near a KK monopole}\label{InstKK}
Applying T-daulity to the primary root and the hook
 of (\ref{dzkk}) we get a
D-instanton inside a KK-monopole.
The primary root $\a_0$ is a D-instanton with action
${1\over \lam}$. The bait, $\g$ is a KK-monopole with action
${1\over {\lam^2}}l_1\cdots l_6 l_7^2$.
The hook $\b$ corresponds to a B-field along the $7^{th}$ and 
$1^{st}$ directions, i.e. a string with action $l_1 l_7$.
Although this system differs from that studied in (\ref{dzkk})
by the sign of $\lrp{\a_0}{\b}$ and $\lrp{\g}{\b}$, it
is likely to have similar features.
The B-field
modifies the D-instanton action and forces
the instanton to sit at the origin.
Let $y$ be a coordinate along the $7^{th}$ circle, and let
us take the $8,9,10$
directions to be noncompact with coordinates $x_8,x_9,x_{10}$
and choose spherical coordinates with $r$ being the distance
to the origin and $\Omega$ being a coordinate on the $\MS{2}$.
The metric of a KK-monopole is the Taub-NUT metric:

\be\label{tnmet}
ds^2 = l_7^2 U(dy - A_i dx^i)^2 + U^{-1} (d\vec{x})^2,\qquad
i=8\dots 10,\qquad 0\le y\le 2\pi.
\ee
where,
$$
U = \left(1 + {{l_7}\over {2|\vec{x}|}}\right)^{-1},
$$
and $A_i$ is the gauge field of a monopole centered at the origin.

The 2-form $B_{\u\v}dx^\u\wdg dx^\v$ has to be proportional to
$dy\wdg dx_1$. However, in the presence of the Taub-NUT metric,
$dy$ is not globally defined over the sphere $\MS{2}$.
Instead, $dy - A_i dx^i$ is the well-defined angular-form.
However, $B= (dy-A_i dx^i)\wdg dx_1$ has a non-vanishing field
strength $H=dB = F\wdg dx_1$, where $F$ is the 2-form
field-strength of a monopole on $\MS{2}$.
The presence of the nonzero $H$ will modify the equations
of motion both for the metric as well as the dilaton.
The dilaton will now have a maximum as $r\rightarrow 0$.
This will make the action ${1\over\lam}$ smaller at the origin.
The difference in the instanton action ${1\over \lam}$ at
infinity and in the core of the instanton should be given exactly
by eqn~(\ref{Ttwo}).
The second derivative of the function ${1\over \lam}$
at the origin will create an effective quadratic term
for the the fields $X^I$ in (\ref{Izero}).
Although the field-strength $H=dB$ could be large at
the origin, we will assume that it has no effect on
the D(-1)-brane action. The exact solution will be explored
further in \cite{WIP}.

Let us also note in passing that 
we can similarly study an M2-brane near a KK-monopole in M-theory.
This time the metric along the directions of the M2-brane
will probably be smaller at the origin which will cause
the M2-brane to be attracted to the center.


\section{Deformations with two-hooks and more}\label{twhook}
So far we have considered examples that deform the integral
in (\ref{Izero}) by a mass term (\ref{Imass}) that preserves
half the supersymmetry and gives mass to 4 out of the 10 $X_I$'s.
Our final goal is to give mass to all the fields and also
break supersymmetry completely.
As a first step we will add a mass term that preserves only
${1\over 4}$ of the SUSY.
We will find it easy to use a model similar to the 
elliptic model of \cite{WitBR}.
(For somewhat related constructions see \cite{LPTi,HU}).

\subsection{Two NS5-branes: variant I}\label{twoNSI}
We start with a D1-brane with Euclidean world-sheet
stretching along directions $1,2$.
we add an NS5-brane along directions $2,3,4,5,6,7$
and add a Dehn twist such that as we go around the $1^{st}$ 
direction we translate along the $3^{rd}$.
We add a second NS5-brane along directions $1,3,4,5,6,8$.
We also add a Dehn twist such that as we go around the $2^{nd}$
direction we translate along the $4^{th}$ direction.
Note that both NS5-branes include the directions $3,4$.
Each NS5-brane creates a mass in the directions orthogonal to it.

The configuration preserves ${1\over 4}$ supersymmetry.
Let us calculate the intersection matrix.
We define the roots corresponding to the branes as follows:
\bear
T_{\a_0} &=& {1\over \lam}l_1 l_2,\nn\\
T_{\b_1} &=& l_1 l_3^{-1},\nn\\
T_{\g_1} &=& {1\over {\lam^2}} l_2 l_3 l_4 l_5 l_6 l_7,\nn\\
T_{\b_2} &=& l_2 l_4^{-1},\nn\\
T_{\g_2} &=& {1\over {\lam^2}} l_1 l_3 l_4 l_5 l_6 l_8,\nn
\eear
They give the corresponding products:

\begin{tabular}{c|rrrr|}        
       & $\g_1$ & $\b_1$ & $\g_2$ & $\b_2$ \\ \hline
$\g_1$ &    2   &   -1   &    0   &    0   \\
$\b_1$ &   -1    &   2   &    0   &    0   \\
$\g_2$ &    0   &    0   &    2   &   -1   \\
$\b_2$ &    0    &   0   &   -1   &    2   \\ \hline
\end{tabular}

The eigenvalues of the mass term in (\ref{Imass}) are proportional to
(according to the rules for ``brane-boxes'' \cite{WitBR,HU}):
\be\label{MelbrI}
M^2\sim
2\left\{ \left({{\phi_{\b_1}}\over {l_3}}\right)^2 \right\},\,
2\left\{ \left({{\phi_{\b_2}}\over {l_4}}\right)^2 \right\},\,
2\left\{ \left({{\phi_{\b_1}}\over {l_3}}\right)^2
        +\left({{\phi_{\b_2}}\over {l_4}}\right)^2
 \right\},\, 4\left\{ 0\right\}.
\ee

\subsection{An NS5-brane and a KK-monopole: variant-I}
Analyzing the constructions with two NS5-branes involves some
guesswork because the dynamics of strings joining
the open ends of the D-branes (and necessarily passing through
NS5-branes) is strongly coupled.

Instead, we will present a U-dual construction which, we believe,
is simpler to analyze.
We start with the elliptic brane configuration
of a D2-brane ending on an NS5-brane.
This system realizes the dimensional reduction of a system
with $\SUSY{2}$ in 3+1D and a massive adjoint hypermultiplet.
We can now immerse that construction inside a KK-monopole
that will realize an R-symmetry twist as we go along the other
direction of the D2-brane.
The corresponding actions are:
\bear
T_{\a_0} &=& {1\over \lam}l_1 l_2,\nn\\
T_{\b_1} &=& l_1 l_3^{-1},\nn\\
T_{\g_1} &=& {1\over {\lam^2}} l_2 l_3 l_4 l_5 l_6 l_7,\nn\\
T_{\b_2} &=& l_2 l_6^{-1},\nn\\
T_{\g_2} &=& {1\over {\lam^2}} l_1 l_2 l_3 l_4 l_5 l_6^2 l_8,\nn
\eear
Note that the NS5-brane is wrapped on a 2-manifold inside
the 4D space transverse to the KK-monopole (the Taub-NUT space).
This 2-manifold includes the Taub-NUT direction and is smooth.
Because there are no new singularities other than those
already present in the elliptic models of \cite{WitBR}, we 
can argue that at low energies the construction gives a
term in (\ref{Imass}) with $M^2$ having the same form as 
(\ref{MelbrI}).
$$
M^2\sim
2\left\{ \left({{\phi_{\b_1}}\over {l_3}}\right)^2 \right\},\,
2\left\{ \phi_{\b_2}^2 \right\},\,
2\left\{ \left({{\phi_{\b_1}}\over {l_1}}\right)^2
+\phi_{\b_2}^2 \right\},\, 4\left\{ 0\right\}.
$$
After T-duality to obtain the previous example, it is easily seen that
this agrees with the rules for ``brane-boxes'' \cite{HU}.
The intersection matrix is as before:

\begin{tabular}{c|rrrr|}        
       & $\g_1$ & $\b_1$ & $\g_2$ & $\b_2$ \\ \hline
$\g_1$ &    2   &   -1   &    0   &    0   \\
$\b_1$ &   -1    &   2   &    0   &    0   \\
$\g_2$ &    0   &    0   &    2   &   -1   \\
$\b_2$ &    0    &   0   &   -1   &    2   \\ \hline
\end{tabular}

\subsection{Two NS5-branes: variant II}\label{twoNSII}
We start with a D1-brane with Euclidean world-sheet
stretching along directions $1,2$.
we add an NS5-brane along directions $2,3,4,5,6,7$
and add a Dehn twist such that as we go around the $1^{st}$ 
direction we translate space along the $3^{rd}$ direction.
We add a second NS5-brane along directions $1,3,4,5,6,8$.
We also add a Dehn twist such that as we go around the $2^{nd}$
direction we translate along the same $3^{rd}$ direction.
Note that both NS5-branes include the direction $3$ as they should.
Each NS5-brane creates a mass in the directions orthogonal to it.
We define the roots corresponding to the branes as follows:
\bear
T_{\a_0} &=& {1\over \lam}l_1 l_2,\nn\\
T_{\b_1} &=& l_1 l_3^{-1},\nn\\
T_{\g_1} &=& {1\over {\lam^2}} l_2 l_3 l_4 l_5 l_6 l_7,\nn\\
T_{\b_2} &=& l_2 l_3^{-1},\nn\\
T_{\g_2} &=& {1\over {\lam^2}} l_1 l_3 l_4 l_5 l_6 l_8,\nn
\eear
They give the corresponding products:

\begin{tabular}{c|rrrr|}        
       & $\g_1$ & $\b_1$ & $\g_2$ & $\b_2$ \\ \hline
$\g_1$ &    2   &   -1   &    0   &    0   \\
$\b_1$ &   -1    &   2   &    0   &    1   \\
$\g_2$ &    0   &    0   &    2   &   -1   \\
$\b_2$ &    0    &   1   &   -1   &    2   \\ \hline
\end{tabular}

This time the masses in (\ref{Imass}) are:

\be\label{NSvarII}
M^2\sim
2\left\{ \left({{\phi_{\b_1}}\over {l_3}}\right)^2 \right\},\,
2\left\{ \left({{\phi_{\b_2}}\over {l_4}}\right)^2 \right\},\,
\left\{ \left({{\phi_{\b_1}}\over {l_3}}
       +{{\phi_{\b_2}}\over {l_4}}\right)^2\right\},\,
\left\{ \left({{\phi_{\b_1}}\over {l_3}}
        -{{\phi_{\b_2}}\over {l_4}}\right)^2\right\},\,
4\left\{ 0\right\}.
\ee
We can replace $T_{\a_0}$ with ${1\over \lam} l_1 l_2 l_4 l_5 l_6$
to obtain a 3D theory and
therefore the instanton integral can be the dimensional
reduction of a supersymmetric 3D theory.
It cannot be the dimensional reduction
of a supersymmetric
4D theory because one of the masses is not doubled.

It is also not completely clear to me if new Yukawa couplings
are generated or not
(unlike the case of variant-I where the KK-monopole derivation
was safe, at least for $l_i\gg 1$ and $\lam\ll 1$).

\subsection{Two KK-monopoles}\label{twoKK}
The above construction is U-dual to the following:
\bear
T_{\a_0} &=& {1\over \lam}l_1 l_2 l_3 l_4,\nn\\
T_{\b_1} &=& l_1 l_7^{-1},\nn\\
T_{\g_1} &=& {1\over {\lam^2}} l_1 l_2 l_3 l_4 l_5 l_6 l_7^2,\nn\\
T_{\b_2} &=& l_1 l_6^{-1},\nn\\
T_{\g_2} &=& {1\over {\lam^2}} l_1 l_2 l_3 l_4 l_7 l_8 l_6^2,\nn
\eear
This system seems to describe the dimensional reduction of 
$\SUSY{4}$ SYM compactified on $\MS{1}$ (the $1^{st}$ direction)
with an $SO(6)$ R-symmetry twist along that direction.
Each pair of hook and bait $(\b_j,\g_j)$ ($j=1,2$)
on its own creates a twist with $SU(4)\sim SO(6)$ eigenvalues:
$$
(e^{i\phi_{\b_j}},\,
e^{-i\phi_{\b_j}},\, 0,\, 0).
$$
However, the masses in (\ref{NSvarII}) cannot be obtained
from the limit of a small twist in $SU(4)$.
If we compactify $\SUSY{4}$ SYM on $\MS{1}$ with an R-symmetry
twist with $SU(4)$ eigenvalues:
$$
(e^{i\a_1},\,
e^{i\a_2},\,
e^{i\a_3},\,
e^{-i(\a_1 + \a_2 + \a_3)}),
$$
Then the bare 3D masses of the scalars are going to be proportional
to:
$$
M^2\sim 
2\left\{|\a_1+\a_2|^2\right\},\,
2\left\{|\a_1+\a_3|^2\right\},\,
2\left\{|\a_2 + \a_3|^2\right\}.
$$
This does not agree with (\ref{NSvarII}).
There is no immediate contradiction, though, because the configuration
of two KK-monopoles cannot be realized geometrically.
Once we compactify the transverse space to one monopole, we
cannot find a solution any more.

\subsection{A deformation with three hooks}\label{three}
Let us consider a combination of a D2-brane and 3 NS5-branes
as follows:
\bear
T_{\a_0} &=& {1\over \lam}l_1 l_2 l_3,\nn\\
T_{\b_1} &=& l_1 l_4^{-1},\nn\\
T_{\g_1} &=& {1\over {\lam^2}} l_2 l_3 l_4 l_5 l_6 l_7,\nn\\
T_{\b_2} &=& l_2 l_5^{-1},\nn\\
T_{\g_2} &=& {1\over {\lam^2}} l_1 l_3 l_4 l_5 l_6 l_8,\nn\\
T_{\b_3} &=& l_3 l_6^{-1},\nn\\
T_{\g_3} &=& {1\over {\lam^2}} l_1 l_2 l_4 l_5 l_6 l_9.\nn
\eear
This is chosen so that the configuration preserves ${1\over {16}}$
of the supersymmetry.
The corresponding deformation in (\ref{Imass})  has $M^2$ with
eigenvalues proportional to:
\bear
M^2 &\sim&
 \left({{\phi_{\b_1}}\over {l_4}}\right)^2,\,
 \left({{\phi_{\b_2}}\over {l_5}}\right)^2,\,
 \left({{\phi_{\b_3}}\over {l_6}}\right)^2,\,
\nn\\ &&
 \left({{\phi_{\b_1}}\over {l_4}}\right)^2
        +\left({{\phi_{\b_2}}\over {l_5}}\right)^2,\,
 \left({{\phi_{\b_1}}\over {l_4}}\right)^2
        +\left({{\phi_{\b_3}}\over {l_6}}\right)^2,\,
 \left({{\phi_{\b_2}}\over {l_5}}\right)^2
        +\left({{\phi_{\b_3}}\over {l_6}}\right)^2,\,
\nn\\ &&
 \left({{\phi_{\b_1}}\over {l_4}}\right)^2
     +\left({{\phi_{\b_2}}\over {l_5}}\right)^2
        +\left({{\phi_{\b_3}}\over {l_6}}\right)^2,\,
 3\left\{ 0\right\}.
\eear


\section{Instanton pairs with $\lrp{\a}{\b}=-2$}\label{mintwo}
Our goal is to add hooks and baits such that the induced D-instanton
integral will have no supersymmetry at all and also will have no
flat directions.
As long as the primary root $\a_0$ and all the hooks and baits
can be realized as particles in M-theory on $\MT{7}$ with a Euclidean
world-line around an extra $\MS{1}$, it is obvious that 
some supersymmetry will be preserved.
This is because for any configuration of charges in M-theory
on $\MT{7}$, one can find the maximal eigenvalue of the central
charge and get a corresponding BPS state.
It is also likely that inside M-theory on $\MT{8}$ we cannot
completely break supersymmetry with a combination
of instantons corresponding to positive roots.

In order to break supersymmetry completely,
it is very likely that we need to go beyond M-theory
on $\MT{8}$ and therefore go beyond the finite group $E_8$.
One of the new features that the infinite groups $E_9$ and
$E_{10}$ have is pairs of roots with $\lrp{\a}{\b}=-2$.
We will see in section (\ref{bmintw}) that adding two baits 
$\g_1$ and $\g_2$
that satisfy $\lrp{\g_1}{\g_2}=-2$ has, on the face of it,
the potential to break supserymmetry, in certain cases.

In this section we will study in more detail
the cases in which the main root $\a_0$ and the bait $\b$ satisfy
$$
\lrp{\a_0}{\b}=-2,\qquad\a_0^2 = \b^2 = 2.
$$
Various U-dual examples are:
\begin{itemize}
\item
A D(-1)-brane (with action ${1\over \lam}$) near a D7-brane
(with action ${1\over\lam}l_1\cdots l_8$) in type-IIB on $\MT{8}$.

\item
A D0-brane near a D8-brane in type-IA.
This system was studied in \cite{BSSil,DFK,BGL}.

\item
A KK-monopole with respect to the $8^{th}$ direction and with
action $R_1\cdots R_7 R_8^2$ and a KK-monopole with respect to
the $9^{th}$ direction with action $R_1\cdots R_7 R_9^2$ in M-theory 
on $\MT{9}$.

\item
A KK-monopole with respect to the $9^{th}$ direction (with action
${1\over {\lam^2}}l_1\cdots l_6 l_9^2$) intersecting a D7-brane
(with action ${1\over \lam}l_1\cdots l_8$) in type-IIB on $\MT{9}$
(formally).

\item
An NS5-brane (with action ${1\over {\lam^2}}l_1\cdots l_6$)
submerged inside a D8-brane (with action ${1\over\lam}l_1\cdots l_9$)
in type-IIA on $\MT{9}$ (formally).

\item
A D7-brane (with action ${1\over \lam}l_1\cdots l_8$)
intersecting a D3-brane (with action ${1\over \lam}l_1 l_2 l_9 l_{10}$)
along a 2-dimensional plane.
\end{itemize}

\subsection{D(-1)-brane near a D7-brane}
In this case the $\a_0$ root corresponds to a D-instanton with
action ${1\over \lam}$ and the bait is $\b$, a D7-brane with action
${1\over \lam}l_1\cdots l_8$.
On their own, the $N$ D-instantons will be described by the action
$I_0$ of (\ref{Izero}).
We wish to know the effect of the D7-brane.
The D7-brane changes the value of the complex dilaton in the space
around it such that the D(-1)-brane action, $e^{-{1\over \lam} + i\chi}$,
becomes $e^{\log (\barz/\Lambda)} = {\barz\over\Lambda}$.
Here, $\Lambda$ is a cutoff which in the usual case of F-theory
signifies the presence of another $(p,q)$ 7-brane at that distance.
$|z|$ is the distance to the origin where the D7-brane is located.
On top of that, there are fermionic variables that come from quantizing
the open strings with one end on the D(-1)-brane
and the other on the D7-brane.
These variables have mass $|z|$.
With a single D(-1)-brane,
they produce a factor of $z$ when integrated (see \cite{GanZ}).
Together they produce a prefactor of $|z|^2$.
This is just as well, since the phase of the coordinate $z$
is arbitrary and depends on our choice of coordinates.

In the case of $N$ D-instantons, the natural generalization
seems to be $|\det (X_9 + i X_{10})|^2$ where  $X_9$ and $X_{10}$
are $N\times N$ matrices. 
The argument for this is that we get a factor of
$\det (X_9 + i X_{10})$ from integrating the fermionic variables.
To cancel the phase, we expect to get the complex conjugate from
the heuristic $\prod_1^N e^{-{1\over \lam_i} + i\chi_i}$
where $\lam_i$ and $\chi_i$ are the dilaton values at
the positions of the D-instantons -- which makes sense only when
they are far apart.
We propose that the modification to the action due to
the D7-brane is a term
$$
I_d = 2 \log |\det (X_9 + i X_{10})|.
$$

\subsection{NS5-brane near a D8-brane}
Another system that falls into the category of the present
discussion is an NS5-brane near a D8-brane.
We realize this system by considering
an NS5-brane in type-IA on $\MS{1}/\BZ_2$ \cite{PolWit}.
We can T-dualize along the segment to obtain
type-I on $\MS{1}$ as in \cite{PolWit} and the NS5-brane would
become a KK-monopole with respect to $\MS{1}$.
The position of the D8-branes in the original
type-IA system is related to the $SO(32)$ 
Wilson line along $\MS{1}$ and the position 
of the NS5-brane is related to a 2-form flux in the Taub-NUT
solution corresponding to the KK-monopole.
\footnote{I am grateful to
S. Sethi for discussions on this system.}
Now take $N$ such KK-monopoles.
There are low-energy fields which classically
come from the $A_{N-1}$ singularity at the core of the solution.
After S-duality the question becomes
what lives on an $A_{N-1}$
singularity in the heterotic string.
This question was recently studied in \cite{SenD,WitADE}.
We will not discuss it further here.

In \cite{GanXi}, we suggested that a partition function
for M-theory on a space $X$ built as an $\MR{2}$ fibration
over $\MT{9}$ might exist.
We proposed that the $SO(2)$ twists in the $MR{2}$, along the
$i^{th}$ direction of $\MT{9}$ might be captured
by the following hook and bait:
$$
T_\b = R_i R_{10}^{-1},\qquad
T_\g = R_1 R_2 R_3 R_4 R_5 R_6 R_7 R_8 R_9 R_{10}^3.
$$
Formally, $\g$ corresponds to a D8-brane if we pick
the $10^{th}$ direction for the M-theory/type-IIA reduction.
If we now insert an M5-brane with:
$$
T_{\a_0} = R_1 R_2 R_3 R_4 R_5 R_6,
$$
we conjecture that for an appropriate limit of all the $R_j$'s,
the phase $\phi_\b$ will be related to a nonsupersymmetric
twist in the partition function of the M5-brane.
If we formally take the transverse directions of the M5-brane
to be $7\dots 11$ then the twist will be in the $SO(2)$ that
rotates directions $10,11$ (because it will have to preserve
directions $7,8,9$.
The intersection matrix of the relevant roots is:

\begin{tabular}{c|rrrr|}
       & $\a_0$ & $\b$ & $\g$ \\ \hline
$\a_0$ &    2   &   1  &  -2  \\
$\b$   &    1   &   2  &  -2  \\
$\g$   &   -2   &  -2  &   2  \\  \hline
\end{tabular}

If the conjecture is true, it seems that the $-2$ products
of roots play a crucial role in supersymmetry breaking.


\section{Pairs of baits with $\lrp{\a}{\b}=-2$}\label{bmintw}
We will now study the effect of having two baits $\g_1$ and $\g_2$
with $\lrp{\g_1}{\g_2}=-2$.
The first example preserves ${1\over 4}$ of the supersymmetry
but is useful for getting rid of many flat directions.
The rest of the examples in this section seem to break supersymmetry
completely.

\subsection{A D-instanton near two KK-monopoles}\label{DtwoKK}
We have conjectured in subsection~(\ref{InstKK})
that $N$ D-instantons near a KK-monopole with an appropriate
NSNS B-field flux at infinity are pinned to the center
of the Taub-NUT space and are described by a mass deformation
(\ref{Imass}) that breaks ${1\over 2}$ supersymmetry.
What happens if we insert another KK-monopole?
Let us take the following actions (written formally for
$\MT{10}$ to indicate the directions):
\bear
T_{\a_0} &=& {1\over \lam},\nn\\
T_{\b_1} &=& l_1 l_7,\nn\\
T_{\g_1} &=& {1\over {\lam^2}} l_1 l_2 l_3 l_4 l_5 l_6 l_7^2,\nn\\
T_{\b_2} &=& l_2 l_6,\nn\\
T_{\g_2} &=& {1\over {\lam^2}} l_1 l_2  l_6^2 l_7 l_8 l_9 l_{10},\nn
\eear
The configuration of the two KK-monopoles is certainly
a solution and can even be
realized as a decompactification limit of type-II on $K_3\times K_3$.
We conjecture that together this configuration 
gives mass to $X_3\dots X_{10}$ and leaves $4$ supersymmetries
in the D-instanton action. The action is a dimensional
reduction of a 2D theory with $\SUSY{(4,0)}$.
The 2D theory is just a $U(N)$ gauge theory with two mass terms.

The intersection matrix is:
\begin{tabular}{c|rrrr|}        
       & $\g_1$ & $\b_1$ & $\g_2$ & $\b_2$ \\ \hline
$\g_1$ &    2   &   -1   &   -2   &    0   \\
$\b_1$ &   -1    &   2   &    0   &    0   \\
$\g_2$ &   -2   &    0   &    2   &   -1   \\
$\b_2$ &    0    &   0   &   -1   &    2   \\ \hline
\end{tabular}

The corresponding mass term in (\ref{Imass}) has eigenvalues
proportional to:
\be\label{twokk}
M^2\sim
4\left\{ \phi_{\b_1}^2 \right\},\,
4\left\{ \phi_{\b_2}^2 \right\},\,
2\left\{ 0\right\}.
\ee
I do not know if there is a solution of $\a_0,\b_1,\b_2,\g_1,\g_2$
inside $E_{10}$ or one actually has to consider $E_{11}$ to 
realize the roots as branes.

We can combine this construction with that of (\ref{twoNSI})
to obtain a system that removes all flat directions:
\be
M^2\sim
2\left\{ \left({{\phi_{\b_1}}\over {l_3}}\right)^2 \right\},\,
2\left\{ \left({{\phi_{\b_2}}\over {l_4}}\right)^2 \right\},\,
2\left\{ \left({{\phi_{\b_1}}\over {l_3}}\right)^2
        +\left({{\phi_{\b_2}}\over {l_4}}\right)^2
 \right\},\, 4\left\{ \phi_{\b_3}^2\right\}
\ee
and we need to find roots with intersection matrix:

\begin{tabular}{c|lrrrrrrrrr|}
       & $\a_0$ & $\g_1$ & $\b_1$ & $\g_2$ & $\b_2$ &
                  $\g_3$ & $\b_3$  \\ \hline
$\a_0$ & 2& 0&-1& 0&-1& 0&-1 \\
$\g_1$ & 0& 2&-1& 0& 0&-2& 0 \\
$\b_1$ &-1&-1& 2& 0& 0& 0& 0 \\
$\g_2$ & 0& 0& 0& 2&-1&-2& 0 \\
$\b_2$ &-1& 0& 0&-1& 2& 0& 0 \\
$\g_3$ & 0&-2& 0&-2& 0& 2&-1 \\
$\b_3$ &-1& 0& 0& 0& 0&-1& 2 \\ \hline
\end{tabular}

\subsection{Supersymmetry breaking twists}\label{sbtwst}
Instead of getting rid of the noncompact moduli by mass terms,
as we did above, one can also get rid of some of the noncompact moduli
by R-symmetry twists.
For example, we can ask what is the partition function of the
the D0-brane with generic
$$
U(1)_L\times U(1)_R\subset SU(2)_L\times SU(2)_R
\sim SO(4)\subset SO(4)\times SO(6)\subset SO(10)
$$
R-symmetry twists along $\MS{1}$ (as in \cite{KKN}).
We conjecture that to realize it we have to take:
\bear
T_{\a_0} &=& {1\over {\lam}} l_1,\nn\\
T_{\b}  &=& l_1 l_7^{-1},\nn\\
T_{\g} &=& {1\over {\lam^2}} l_1 l_2 l_3 l_4 l_5 l_6 l_7^2,\nn\\
T_{\b'} &=& l_1 l_8^{-1},\nn\\
T_{\g'} &=& {1\over {\lam^2}} l_1 l_2 l_3 l_4 l_5 l_6 l_8^2,\nn
\eear
Here, $\b$ and $\g$ correspond to the
$U(1)_L$ twist and $\b'$ and $\g'$ correspond to the $U(1)_R$
twist.
The intersection matrix is:

\begin{tabular}{c|rrrrr|}
       & $\a_0$ & $\g$ & $\b$ & $\g'$ & $\b'$ \\ \hline
$\a_0$ &    2   &   0  &   1  &   0   &   1   \\
$\g$   &    0   &   2  &  -1  &  -2   &   1   \\
$\b$   &    1   &  -1  &   2  &   1   &   1   \\
$\g'$  &    0   &  -2  &   1  &   2   &  -1   \\
$\b'$  &    1   &   1  &   1  &  -1   &   2   \\ \hline
\end{tabular}

Formally, once the $7^{th}$ direction is compact there is no
Taub-NUT solution with respect to the $8^{th}$ circle.
As we mentioned before, we treat the Taub-NUT solution
only as a motivation for an abstract procedure inside $E_9$ or
$E_{10}$.
However, we have to add a caveat.
We have seen in subsection (\ref{twoKK}) that treating
to intersecting KK-monopoles formally does not always give
the expected intuitive results.
Although I do not have a convincing argument that $\g'$ indeed
does the trick, let us describe another system with 
two baits with $\lrp{\g_1}{\g_2}=-2$ and the following property.
If the $U(1)_L$ twists $\phi_{\b_1} = \cdots =\phi_{\b_6} = 0$
or the $U(1)_R$ twists $\phi_{\b_1'} = \cdots \phi_{\b_6'} = 0$,
then ${1\over 8}$ of the supersymmetry is preserved.
However,
if both twists are turned on then supersymmetry is broken.

\subsection{Interfering hooks: D-instanton and two D3-branes}
This system is made of a D-instanton in the presence of two transverse
Euclidean D3-branes:
\bear
T_{\a_0} &=& {1\over {\lam}},\nn\\
T_{\g} &=& {1\over {\lam}} l_1 l_2 l_3 l_4,\nn\\
T_{\g'} &=& {1\over {\lam}} l_5 l_6 l_7 l_8,\nn
\eear
We will also exhibit two hooks as NSNS B-field fluxes:
\bear
T_{\b} &=& l_1 l_2,\nn\\
T_{\b'} &=& l_5 l_6.\nn
\eear
The intersection matrix is:

\begin{tabular}{c|rrrrr|}
       & $\a_0$ & $\g$ & $\b$ & $\g'$ & $\b'$ \\ \hline
$\a_0$ &    2   &   0  &  -1  &   0   &  -1   \\
$\g$   &    0   &   2  &   1  &  -2   &  -1   \\
$\b$   &   -1   &   1  &   2  &  -1   &   0   \\
$\g'$  &    0   &  -2  &  -1  &   2   &   1   \\
$\b'$  &   -1   &  -1  &   0  &   1   &   2   \\ \hline
\end{tabular}

Note that $\lrp{\a_0}{\b}=-1$ and not $+1$ as before.
Nevertheless, we can see the ``interference'' of the two hooks as
follows.
In the presence of the fluxes the 
D-instanton becomes a noncommutative  Yang-Mills instanton 
inside the D3-brane. However, it can only be a large noncommutative
instanton in one of the D3-branes but not the other.
In the presence of fluxes on both D3-branes, supersymmetry has
to be broken.
Note that $\lrp{\g}{\g'}=-2$.

\subsection{Interfering hooks: surfaces in $\MT{8}$}\label{surfT}
The following system has a related, though somewhat different
behavior.
Here, if any of the hooks is nonzero supersymmetry seems to be
broken, while if both are zero, supersymmetry is preserved.

Take 3 Euclidean D3-branes inside $\MT{8}$
as follows:
\bear
T_{\a_0} &=& {1\over \lam} l_1 l_3 l_5 l_7,\nn\\
T_{\g_1} &=& {1\over \lam} l_2 l_3 l_6 l_7,\nn\\
T_{\g_2} &=& {1\over \lam} l_1 l_4 l_5 l_8,\nn\\
T_{\b_1} &=& l_1 l_2^{-1},\nn\\
T_{\b_2} &=& l_3 l_4^{-1}.\nn
\eear
The intersection matrix is:

\begin{tabular}{c|rrrrr|}
       & $\a_0$ & $\g_1$ & $\b_1$ & $\g_2$ & $\b_2$ \\ \hline
$\a_0$ &    2   &   0    &   1    &   0    &   1   \\
$\g_1$ &    0   &   2    &  -1    &  -2    &   1   \\
$\b_1$ &    1   &  -1    &   2    &   1    &   0   \\
$\g_2$ &    0   &  -2    &   1    &   2    &  -1   \\
$\b_2$ &    1   &   1    &   0    &  -1    &   2   \\ \hline
\end{tabular}

This system describes a (2-complex dimensional) surface inside a product
of two slanted $\MT{4}$'s. Let us denote the first $\MT{4}$ by $X$ and
the second by $Y$. The hooks $\b_1$ and $\b_2$ specify two Dehn twists
in $X$ and $Y$ respectively. We would like to argue that with any of the
two Dehn twists turned on, there is no complex structure on $\MT{8}$
such that the sum of the cohomology classes of the three D3-branes is
analytic  (i.e. a 4-form of type $(2,2)$ in Dolbeaux cohomology).

To begin, let us recall some facts about abelian tori (see \cite{GriHar}).
We can regard $\MT{2n}$ as $\MC{n}/\Lambda$ where $\Lambda$ is a lattice
and we can pick a basis for the lattice $\he_1,\dots,\he_{2n}\in\MC{n}$.
We can also pick a basis of $\MC{n}$ such that the first
$n$ vectors $\he_1\dots,\he_n$ will be unit vectors in $\MC{n}$.
The remaining vectors $\he_{n+1}\dots,\he_{2n}$ form an $n\times n$
matrix $Z$. An {\em abelian variety} is a torus that can be embedded
inside some $\CP{k}$ (for large enough $k$). It can be shown \cite{GriHar}
that $\MT{2n}$ is an abelian variety if and only if one can choose $Z$
to be symmetric and such that $\Im  Z$ is positive definite.

Let us now take the example of $\MT{4}$ constructed as a $\MT{2}$ fibration
over a base $\MT{2}$ with a Dehn twist turned on.
We can pick a coordinate $z_1$ for the base and $z_2$ for the fiber
and we have the identifications:
$$
(z_1,z_2)\sim (z_1+1,z_2)\sim (z_1+\tau',z_2+\lambda')
\sim (z_1,z_2+1)\sim (z_1,z_2+\sigma').
$$
Here $\tau'=\tau'_1 + i\tau'_2$ and
$\sigma'=\sigma'_1 + i\sigma'_2$
are the complex structures of the base and fiber
and $\lambda'$ is the Dehn twist.
Let us change coordinates to:
\bear
w_1 &=& z_1 -{1\over 2}i b(z_1-\bz_1) 
     -{i\over 2}a (z_2 - \bz_2),\nn\\
w_2 &=& z_2 -{1\over 2}i c(z_2-\bz_2) 
     -{i\over 2}a (z_1 - \bz_1),\nn
\eear
where $a,b,c$ are real. This preserves the K\"ahler class 
$$
dw_1\wdg d\bw_1 + dw_2\wdg d\bw_2 = 
dz_1\wdg d\bz_1 + dz_2\wdg d\bz_2.
$$
The matrix $Z$ takes the form:
$$
\left(\begin{array}{cc}
\tau + b\tau'_2 + a\lam'_2 & a \sigma'_2 \\
\lambda' + c\lam'_2 + a\tau'_2 & \sigma' + c\sigma'_2 \\
\end{array}\right)
$$

Let us denote the 6 cycles in the integer homology group $H_2(\MT{4},\BZ)$
as $\lbrack \he_i\he_j\rbrack$ ($1\le i<j\le 4$).
Let us look for a $(1,1)$ form, $\omega$, 
such that:
$$
\int_{\lbrack \he_1\he_3\rbrack} \omega = n,\qquad
\int_{\lbrack \he_2\he_4\rbrack} \omega = k,
$$
and all the other $\int_{\lbrack \he_2\he_4\rbrack} \omega = 0$.
It is easy to check that such an $\omega$ exists if and only if
$Z_{12} = {k\over n} Z_{21}$.

Now we return to $\MT{8}$ and take $Z$ to be of the form:
$$
Z=\left(\begin{array}{cccc}
\tau & \lam & 0 & 0 \\
\lam & \sig & 0 & 0 \\
0 & 0 & \wtau & \wlam \\
0 & 0 & \wlam & \wsig \\
\end{array}\right)
$$
We wish to find a $(2,2)$ form, $\omega$ on $\MT{8}=\MT{4}\times \MT{4}$
such that
$$
\int_{\lbrack\he_1\he_3\he_5\he_7\rbrack}\omega =n,\qquad
\int_{\lbrack\he_1\he_4\he_5\he_8\rbrack}\omega =
\int_{\lbrack\he_2\he_3\he_6\he_7\rbrack}\omega =1,
$$
and all the other $\int_{\lbrack\he_i\he_j\he_k\he_l\rbrack}\omega =0$.
There are $70$ 4-cycles $\lbrack\he_i\he_j\he_k\he_l\rbrack$.
We get 70 equations in 70 variables. The coefficients are the $4\times 4$
minors of the matrix:
$$
C = 
\left(\begin{array}{cccccccc}
1 & 0 & 0 & 0 & \tau & \lam & 0 & 0 \\
0 & 1 & 0 & 0 & \lam & \sig & 0 & 0 \\
0 & 0 & 1 & 0 & 0 & 0 & \wtau & \wlam \\
0 & 0 & 0 & 1 & 0 & 0 & \wlam & \wsig \\
1 & 0 & 0 & 0 & \btau & \blam & 0 & 0 \\
0 & 1 & 0 & 0 & \blam & \bsig & 0 & 0 \\
0 & 0 & 1 & 0 & 0 & 0 & \bwtau & \bwlam \\
0 & 0 & 0 & 1 & 0 & 0 & \bwlam & \bwsig \\
\end{array}\right).
$$
The coefficient matrix is $C^{(4)}$ and the inverse matrix is proportional
to the complementary minors.
Up to a constant, we find the following 4-form that is Poincar\`e dual
to the homology class:
\bear
\omega &=& {1\over \Delta}(n\omega_1 + \omega_2 + \omega_3),\nn\\
\Delta &\equiv& {1\over 4}\det C =
4 (\lam_2^2 -\tau_2\sig)
(\wlam_2^2 -\wtau_2\wsig_2),\nn
\eear
\bear
\omega_1 &=&
-4\tau_2\wtau_2 w_1\wdg w_3\wdg\bw_1\wdg\bw_3
\nn\\ &&
-2i\tau_2\bwlam w_1\wdg w_3\wdg\bw_1\wdg\bw_4
-2i\wtau_2\blam w_1\wdg w_3\wdg\bw_2\wdg\bw_3
\nn\\ &&
+2i\tau_2\wlam w_1\wdg w_4\wdg\bw_1\wdg\bw_3
+2i\wtau_2\lam w_2\wdg w_3\wdg\bw_1\wdg\bw_3
\nn\\ &&
+2i\wtau_2\lam w_1\wdg w_2\wdg w_3\wdg\bw_3
+2i\tau_2\wlam w_1\wdg w_3\wdg w_4\wdg\bw_1
\nn\\ &&
+2i\tau_2\bwlam w_1\wdg\bw_1\wdg\bw_3\wdg\bw_4
+2i\wtau_2\blam w_3\wdg\bw_1\wdg\bw_2\wdg\bw_3
+O(\lam)^2
\nn
\eear
\bear
\omega_2 &=&
-4\tau_2\wsig_2 w_1\wdg w_4\wdg\bw_1\wdg\bw_4
\nn\\ &&
+2i\tau_2\wlam w_1\wdg w_3\wdg\bw_1\wdg\bw_4
-2i\tau_2\bwlam w_1\wdg w_4\wdg\bw_1\wdg\bw_3
\nn\\ &&
-2i\wsig_2\blam w_1\wdg w_4\wdg\bw_2\wdg\bw_4
+2i\wsig_2\lam w_2\wdg w_4\wdg\bw_1\wdg\bw_4
\nn\\ &&
+2i\wsig_2\lam w_1\wdg w_2\wdg w_4\wdg\bw_4
-2i\tau_2\wlam w_1\wdg w_3\wdg w_4\wdg\bw_1
\nn\\ &&
-2i\tau_2\bwlam w_1\wdg\bw_1\wdg\bw_3\wdg\bw_4
+2i\wsig_2\blam w_4\wdg\bw_1\wdg\bw_2\wdg\bw_4
\nn\\ &&
+O(\lam)^2
\nn
\eear
\bear
\omega_3 &=&
-4\sig_2\wtau_2 w_2\wdg w_3\wdg\bw_2\wdg\bw_3
\nn\\ &&
+2i\wtau_2\lam w_1\wdg w_3\wdg\bw_2\wdg\bw_3
-2i\wtau_2\blam w_2\wdg w_3\wdg\bw_1\wdg\bw_3
\nn\\ &&
-2i\sig_2\bwlam w_2\wdg w_3\wdg\bw_2\wdg\bw_4
+2i\sig_2\wlam w_2\wdg w_4\wdg\bw_2\wdg\bw_3
\nn\\ &&
-2i\wtau_2\lam w_1\wdg w_2\wdg w_3\wdg\bw_3
+2i\sig_2\wlam w_2\wdg w_3\wdg w_4\wdg\bw_2
\nn\\ &&
-2i\wtau_2\blam w_3\wdg\bw_1\wdg\bw_2\wdg\bw_3
+2i\sig_2\bwlam w_2\wdg\bw_2\wdg\bw_3\wdg\bw_4
\nn\\ &&
+O(\lam)^2
\nn
\eear
It is easy to check that there is no $SO(8)$ matrix that when
acting on $w_1,\dots,w_4,\bw_1,\dots,\bw_4$ 
(preserving the metric) brings $\omega$ to a $(4,4)$ form
(at least to first order in $\lam$ and $\wlam$).


\section{Higher dimensional gauge theories}\label{higherd}
So far we focussed on the 0D D-instanton actions.
In this section we will describe various deformations of higher
dimensional gauge theories and the corresponding hooks and baits
that realize them.

\subsection{Mass deformed $\SUSY{4}$ SYM}
In subsection (\ref{ellip}) we described a deformation
that corresponds to a mass term in (\ref{Imass})
that preserves half the supersymmetry.
The same roots $\a_0$,$\b$ and $\g$, in another limiting region
of the parameters $\lam,l_1,\dots,l_7$ can describe a mass
deformation of $\SUSY{4}$ SYM exactly as in \cite{WitBR}.
We can take:
$$
T_{\a_0} = {1\over \lam} l_1 l_2 l_3 l_4 l_5,\,
T_{\g} = {1\over {\lam^2}} l_1 l_2 l_3 l_4 l_6 l_7,\,
T_{\b} = l_5 l_6^{-1}.
$$
Similarly, we can get the twisted $(2,0)$ and little-string
theories as in \cite{CGKM}:
$$
T_{\a_0} = {1\over \lam} l_1 l_2 l_3 l_4 l_5 l_6,\,
T_{\g} = {1\over {\lam^2}} l_1 l_2 l_3 l_4 l_5 l_6 l_7^2,\,
T_{\b} = l_6 l_7^{-1}.
$$
These constructions preserve ${1\over 2}$ of the supersymmetry
and we conjecture that to break supersymmetry we need
to add more hooks and baits.
For example, one can add more R-symmetry twists.
We can ask what is the partition function of the
little-string theory with generic
$$
U(1)_L\times U(1)_R\subset SU(2)_L\times SU(2)_R
$$
R-symmetry twists along $\MT{6}$.
This question was raised in \cite{GanXi}.
To get the answer out of $\ZM$ we take
$$
T_{\a_0} = {1\over {\lam^2}} l_1 l_2 l_3 l_4 l_5 l_6.
$$
We propose to add hooks 
$$
T_{\b_i} = l_i l_7^{-1}
$$
corresponding to the
$U(1)_L$ twists and a corresponding KK-monopole bait:
$$
T_{\g} = {1\over {\lam^2}} l_1 l_2 l_3 l_4 l_5 l_6 l_7^2.
$$
In order to trap the other $U(1)_R$ twists we conjecture that we need
to add hooks:
$$
T_{\b_i'} = l_i l_8^{-1}
$$
and a second bait:
$$
T_{\g'} = {1\over {\lam^2}} l_1 l_2 l_3 l_4 l_5 l_6 l_8^2.
$$
As discussed in subsection (\ref{sbtwst}),
once the $7^{th}$ direction is compact there is no
Taub-NUT solution with respect to the $8^{th}$ circle.
However, as we mentioned before, we treat the Taub-NUT solution
only as a motivation for an abstract procedure inside $E_{10}$.

\subsection{The particle spectrum}\label{parspec}
In principle we can also ``see'' (at least part of) the spectrum
by Fourier transforming with respect to appropriate phases.
Let us look again at the elliptic brane configuration of \cite{WitBR}
in 3+1D.
We have:
\bear
T_{\a_0} &=& {1\over \lam} l_1 l_2 l_3 l_4 l_5,\nn\\
T_{\b} &=&  l_5 l_6^{-1},\nn\\
T_{\g} &=& {1\over {\lam^2}} l_1 l_2 l_3 l_4 l_6 l_7,\nn
\eear
Suppose we managed to somehow break supersymmetry and get
rid of the remaining moduli by adding more, unspecified, hooks
and baits but let us assume that they are small.
Part of the spectrum of the model has an adjoint hypermultiplet
of bare mass proportional to $m=l_6\phi_\b$.
Let us think of $l_1$ as the time direction.
How do we see that the action indeed has contributions of the form
$e^{-m l_1}$?
We can trap the contribution with a given (net) number of such
particles by counting the string winding-number in direction $l_5$.
Note that as far as M-theory is concerned, the string winding number
along the $5^{th}$ direction is an integer.
Even though the massive hypermultiplets come from strings that
seem to have fractional winding number along the $6^{th}$ direction
(a fraction of $\phi_{\b}$) the endpoints of the strings of the D4-brane
are a pair of oppositely charged points and the electric flux
emanating from them along the $5^{th}$ direction effectively
``closes'' the open string.
 Thus, it follows that we need to add a root with corresponding action:
$$
T_{\delta} = l_1 l_5,
$$
and look for terms proportional to $e^{2\pi i k\phi_{\delta}}$ in
$\ZM$ in order to extract the contribution with $k$ massive
hypermultiplets. If we do not Fourier-transform with respect
to $\phi_\delta$, we have to set $\phi_\delta=0$ effectively summing
over all $k$.
Let us also add a fifth root, $\eta$, that corresponds to momentum,
say, around the $4^{th}$ direction.
We take: $T_\eta = l_1 l_4^{-1}$.
Now we expect the behavior:
\be\label{TzkT}
e^{-2\pi (T_0 +k T)
  + 2\pi i(\phi_{\a_0} + \phi_\g +k\phi_\delta +l\phi_\eta)},\qquad
T\equiv \sqrt{l^2 T_\eta^2 
  +k^2\left({{T_\delta}\over {T_\b}}\right)^2\phi_\b^2}.
\ee

Similarly, in the example of the previous subsection which
invloves the twisted little string theory,
we can take $\delta$ such that $T_\delta = l_1 l_7^{-1}$
where $l_1$ is taken as the ``time'' direction.
This corresponds to momentum along the $7^{th}$ circle
which measures R-symmetry charge.
In any case, the relevant roots have intersection matrix:

\begin{tabular}{c|rrrrr|}
         & $\a_0$ &  $\b$  &  $\g$  & $\delta$ & $\eta$ \\ \hline
$\a_0$   &    2   &    1   &    0   &    1     &    0   \\
$\b$     &    1   &    2   &   -1   &    1     &    0   \\
$\g$     &    0   &   -1   &    2   &   -1     &    0   \\
$\delta$ &    1   &    1   &   -1   &    2     &    1   \\
$\eta$   &    0   &    0   &    0   &    1     &    2   \\ \hline
\end{tabular}

It would be interesting to see whether the expected behavior
(\ref{TzkT}) is originating from a harmonic function.

Let us note that one can realize the same intersection matrix
of five roots in M-theory on $\MT{6}$ as follows:
\bear
T_{\a_0} &=& R_1 R_5^{-1},\nn\\
T_\b     &=& R_1 R_4^{-1},\nn\\
T_\g     &=& R_2 R_3 R_4,\nn\\
T_\delta &=& R_1 R_3^{-1},\nn\\
T_\eta   &=& R_2 R_3^{-1}.\label{atoeta}
\eear
This configuration is a ${1\over 8}$-BPS instanton and is likely
to contribute to $\lam^{28}$-terms in the 6D low-energy effective
action of M-theory on $\MT{5}$.

\subsection{The bait for gravity}
In \cite{GanXi} we discussed M-theory on $\MT{7}$ with generic 
$Spin(4)$ twists of the transverse $\MR{4}$.
The twists mean that the
space is an $\MR{4}$ fibration over $\MT{7}$ and
as we go around 1-cycles of $\MT{7}$ we have to rotate the transverse
$\MR{7}$ by an appropriate element of (the spin cover of) $SO(4)$.
We then generalized this construction to include U-duals of twists
but we will not discuss that here.
We proposed that for generic twists there exists a well-defined
partition function of M-theory on this space.
This partition function should be encoded in $\ZM$.

We therefore search for the corresponding hooks and baits.
The natural guess is that for one of the $SU(2)$ factors
we take the bait $\g$ with action $R_1\cdots R_7 R_8^2$ and
hook $\b_i$ with action $R_i R_8^{-1}$ ($i=1\dots 7$) and for the other
$SU(2)$ factor we take the bait $\g'$ with action
$R_1\cdots R_7 R_9^2$ and hook $\b'_i$ with action $R_i R_9^{-1}$.

The product matrix is:
\begin{tabular}{c|cccc|}
        & $\g$ & $\b_j$ &  $\g'$ & $\b'_j$ \\ \hline
$\g$    & $2$  &   $-1$ &   $-2$  &    $1$   \\
$\b_i$  & $-1$ & $2-\delta_{ij}$ &   $1$   &    $\delta_{ij}$   \\
$\g'$   & $-2$ &   $-2$   &   $2$   &   $-1$   \\
$\b'_i$ & $1$  & $\delta_{ij}$ &   $-1$   &    $2-\delta_{ij}$   \\ \hline
\end{tabular}

The system with the two KK-monopoles is hard to analyze because
a semi-classical description of the system is not known.
If we start with a KK-monopole with respect to
the $8^{th}$ direction that
fills directions $1\dots 7$ and compactify another
transverse direction, say the $9^{th}$, it is not clear how
to construct the solution.

Note that in \cite{GanXi}, we suggested a different prescription
for calculating the twisted M-theory action on $\MT{7}$
by starting with the twisted M-theory action on $\MT{9}$.
The present prescription seems to be different but more symmetrical.
I do not know if the two prescriptions agree or not.
Both prescrtiptions are, of course, conjectures.


\section{Harmonic functions
on $E_{10}(\BZ)\backslash E_{10}(\BR)/\Kom$}\label{groups}
In this section we will write down an eqaution for the Laplacian
in terms of the group elements and study some of its properties.

\subsection{$\Nil\Abl\Kom$ decomposition}
We will deal with maximally split Lie group $G$ that can be decomposed
into a product of a nilpotent ($\Nil$), an abelian ($\Abl$) and
a compact ($\Kom$) subgroups.
We take the $\Nil\Abl\Kom$ decomposition to be as follows.
For $\lam\in\Delta$ (the root lattice),
Let $V_\lam$ be the space of elements in 
the Lie algebra with weight $\lam$. $V_0$ is the Cartan subalgebra.
\be
\Nil = e^{\sum\phi_u\tau^u},\qquad
\Abl = e^{\sum\lam_i\tau^i},\qquad
\Kom = e^{\sum c_u(\tau^u - \omega(\tau^u))}.
\ee
Here $\tau^i\in V_0$  and $\tau^u\in V_{\a(u)}$ with $\a(u)\in\Delta_{+}$
a positive root. $\omega$ is the Chevalley involution and in particular
$\omega(\tau^u)\in V_{-\a(u)}$ (i.e. $\omega(\tau^u)$ corresponds
to a negative root).

We also use the Clebsch-Gordan coefficients:
$$
\com{\tau^u}{\tau^v} = \sum_w C^w_{uv}\tau^w.
$$
Here $C^w_{uv}$ are integers.
They are zero unless $\a(w)=\a(u)+\a(v)$.

\subsection{The Laplacian}
The Laplacian is defined to be the quadratic G-invariant operator
of the form:
\bear
\nabla &=& {1\over 2}
   \sum h_{ij} {{\partial^2}\over {\partial\lam_i\partial\lam_j}}
-\sum_i {\partial\over \partial {\lam_i}}
 +\sum W_{u v} e^{\lrp{\lam}{\a(u)}+\lrp{\lam}{\a(v)}}
{{\partial^2}\over {\partial\phi_u \partial\phi_v}}
 +\sum W_{u} e^{\lrp{\lam}{\a(u)}}
{{\partial}\over {\partial\phi_u}}.
\nn\\
&& \label{nabuv}
\eear
Here $\lam_i$ correspond to the simple roots $\a_i\in\Delta_{+}$.
$h_{ij}$ is the Cartan matrix.
The term $\sum_i {\partial/ \partial {\lam_i}}$ can be written
as $\lrp{\delta}{\partial/\partial\lam}$ where $\delta$ is 
half the sum of all positive roots. Although $\delta$ itself
is infinite, the functional $\lrp{\delta}{\cdot}$ is finite and
is given by the formula above.
The functions $W_u$ and $W_{u v}$ are determined by
invariance under the group action and by the requirement
that when all the phases $\phi_w$ are set to zero:
$$
W_{u v}(\{\phi_w=0\}) = \delta_{u v},\qquad
W_u(\{\phi_w=0\}) = 0.
$$
The result is as follows. $W_{u v}$ and $W_u$ are functions
only of
$$
\xi_w\equiv e^{-\lrp{\a(w)}{\lam}}\phi_w.
$$
They satisfy
\bear
0 &=&
\pypx{W_{u v}}{\xi_w}
+\sum_{k=1}^\infty \sum_{u_1,\dots,u_k} \sum_{v_2,\dots,v_k}
b_k \xi_{u_1} \cdots  \xi_{u_n}
C^x_{u_n v_n}\cdots C^{v_3}_{u_2 v_2} C^{v_2}_{u_1 w}
\pypx{W_{u v}}{\xi_x}
\nn\\ &&
-\sum_{k=1}^\infty \sum_{u_1,\dots,u_k} \sum_{v_2\dots,v_k}
\sum_{j=1}^k b_k W_{u u_j}
   \xi_{u_1} \cdots\hat{\xi}_{u_j}\cdots  \xi_{u_n}
C^v_{u_n v_n}\cdots C^{v_3}_{u_2 v_2} C^{v_2}_{u_1 w}
\nn\\ &&
-\sum_{k=1}^\infty \sum_{u_1,\dots,u_k} \sum_{v_2\dots,v_k}
\sum_{j=1}^k b_k W_{v u_j}
   \xi_{u_1} \cdots\hat{\xi}_{u_j}\cdots  \xi_{u_n}
C^u_{u_n v_n}\cdots C^{v_3}_{u_2 v_2} C^{v_2}_{u_1 w}
\label{Wuvxi}\\
0 &=&
\pypx{W_u}{\xi_w}
+\sum_{k=1}^\infty \sum_{u_1,\dots,u_k} \sum_{v_2,\dots,v_k}
b_k \xi_{u_1} \cdots  \xi_{u_n}
C^x_{u_n v_n}\cdots C^{v_3}_{u_2 v_2} C^{v_2}_{u_1 w}
\pypx{W_u}{\xi_x}
\nn\\ &&
-\sum_{k=1}^\infty \sum_{u_1,\dots,u_k} \sum_{v_2\dots,v_k}
  \sum_{j=1}^k b_k W_{u_j}
   \xi_{u_1} \cdots\hat{\xi}_{u_j}\cdots  \xi_{u_n}
C^u_{u_n v_n}\cdots C^{v_3}_{u_2 v_2} C^{v_2}_{u_1 w}
\nn\\ &&
-\sum_{k=1}^\infty \sum_{u_1,\dots,u_k} \sum_{v_2\dots,v_k}
\sum_{j<l=2}^k b_k W_{u_l u_j}
   \xi_{u_1} \cdots\hat{\xi}_{u_j}\cdots\hat{\xi}_{u_l}\cdots\xi_{u_n}
C^u_{u_n v_n}\cdots C^{v_3}_{u_2 v_2} C^{v_2}_{u_1 w}.
\nn\\
&&\label{Wuxi}
\eear
where $b_k$ are the coefficients of
$$
{x\over {1-e^{-x}}} =
\sum_{k=0}^\infty b_k x^k = 1 + {1\over 2}x + {1\over {12}}x^2
 -{1\over {720}}x^4 + {1\over {30240}}x^6 + \cdots
$$
and $\hat{\xi}_{u_j}$ means that the term $\xi_{u_j}$ should
be excluded from the monomial.
One can solve (\ref{Wuvxi}-\ref{Wuxi}) as a power series in $\xi$.
We can start with:
$$
W_{uv} = \delta_{uv} + O(\xi),\qquad W_u = O(\xi).
$$
It is easy to see that $W_u$ and
 $W_{uv}$ will depend only on those $\xi_w$'s that
satisfy either $\a(w)<\a(u)$ or $\a(w)<\a(v)$.
Since the number of positive roots that are smaller than any given
root is finite, it also follows from the iterative procedure
and (\ref{Wuvxi}-\ref{Wuxi}) that $W_{uv}$ and $W_u$ are polynomials
in the $\xi_w$'s.

Now we can pick a positive root $u_0$ and
look for solutions of $\nabla\Phi = 0$
of the form:
$$
\Phi \equiv\Phi(\{\lam_i\},\{\phi_u\}_{\a(u)\le\a(u_0)}).
$$
Since there is only a finite number of $u$'s such that
$\a(u)<\a(u_0)$, and since $W_u$ and $W_{uv}$ are independent
of $\xi_w$'s that do not satisfy $\a(w)<\a(u_0)$,
the equation $\nabla\Phi = 0$ will reduce to a differential equation
in a finite number of variables.
This might be a good approximation for $\ZM$ in regions of
the $\{\lam_i\}$ parameter space that satisfy:
$$
1\ll\, \lrp{\lam}{\b},\qquad {\mbox{if $\b\not{\!\le}\a(u_0)$}}.
$$

\subsection{First order iterative solution}
To first order we find:
\bear
W_{u v} &=& \delta_{uv}+ {1\over 2}\sum_w(C^v_{u w}+C^u_{v w})\xi_w
\nn\\ &&
-{1\over {24}}\sum_{x,y,w}(C^y_{ux} C^v_{yw} +C^y_{uw} C^v_{xy}
  +C^y_{vx} C^u_{yw} +C^y_{vw} C^u_{xy})\xi_x\xi_w
\nn\\ &&
+{1\over 8}\sum_{x,y,w}(C^u_{yx} C^v_{yw} +C^v_{yx} C^u_{yw})\xi_x\xi_w
+O(\xi)^3,
\nn\\
W_u &=& \sum_{x,y} C^u_{xy} C^{y}_{x w}\xi_w +O(\xi)^2
\nn
\eear
The linear term in $W_u$ is nonzero only if $2\a(x)+\a(w) = \a(u)$.
The linear term in $W_{uv}$ is zero unless $\a(w) = \pm(\a(u)-\a(v))$.
The quadratic term with the ${1\over {24}}$ perefactor
is zero unless $\a(x)+\a(w) = \pm(\a(u)-\a(v))$.
The quadratic term with the ${1\over {8}}$ perefactor
is zero unless
$$
\a(w) = \a(v) -\b,\qquad \a(x) = \a(u)-\b,\qquad \b=\a(y)>0,
$$
or:
$$
\a(x) = \a(v) -\b,\qquad \a(w) = \a(u)-\b,\qquad \b=\a(y)>0.
$$
We see that for given
$u,v$ there are only a finite number of terms in the sum.

\subsection{Harmonic functions}
We can now check the statements made in previous sections about
the relation between harmonic functions and actions of branes.

We will start with the examples in section (\ref{hooks}).
First let us take a single BPS instanton.
To make things simple, let us assume that it corresponds to
a simple root $\a_i$.
 From the discussion above it follows that we can look
for a harmonic function $\Phi$
that depends only on $\phi\equiv \phi_{\a_k}$
and no other phases.
Laplace's equation becomes:
$$
{1\over 2}\sum h_{ij}{{\partial^2\Phi}\over {\partial\lam_i\partial\lam_j}}
-\sum {{\partial\Phi}\over {\partial\lam_i}}
+e^{2\lam_k} {{\partial^2\Phi}\over {\partial\phi^2}} = 0.
$$
We are looking for a solution of the form:
$\Phi = e^{-2\pi n T(\lam) +2\pi i n\phi}$.
The function $T$ satisfies:
$$
\pi n\sum h_{ij}{{\partial^2 T}\over {\partial\lam_i\partial\lam_j}}
+2\pi^2 n^2 \sum h_{ij}{{\partial T}\over {\partial\lam_i}}
                      {{\partial T}\over {\partial\lam_j}}
-2\pi n\sum {{\partial T}\over {\partial\lam_i}}
 = 4\pi^2 n^2 e^{2\lam_k}.
$$
We see that $T=e^{\lam_k}$ is a good solution.

Similarly, for pairs of distinct simple roots $\a_k$ and $\a_l$
with $\lrp{\a_k}{\a_l}=0$, one can separate variables and see that
$$
e^{-2\pi (n e^{\lam_k} + m e^{\lam_l}) + 2\pi i(n\phi_k + m\phi_l)},
$$
where $\phi_k\equiv \phi_{\a_k}$ and $\phi_l\equiv\phi_{\a_l}$,
is also a solution.

Now let us take the case $\lrp{\a}{\b}=1$.
In this case $\a-\b$ is also a root and we cannot take both $\a$
and $\b$ to be simple roots.
We can take $\a=\a_k$ and $\b=\a_k+\a_l$ such that $\lrp{\a_k}{\a_l}=-1$.
Now the solution must depend on $\phi_\a\equiv \phi_{\a_k}$
and $\phi_\b$ but also on $\phi_l\equiv\phi_{\a_l}$.
If we take $u$ to be the generator such that $\a(u)=\a_k$,
$v$ the generator such that $\a(v)=\a_l$ and $w$ the generator
such that $\a(w)=\a_k+\a_l$, we find from (\ref{Wuvxi}-\ref{Wuxi})
the equation:
\bear
0 &=& 
{1\over 2}\sum h_{ij}{{\partial^2\Phi}\over {\partial\lam_i\partial\lam_j}}
-\sum {{\partial\Phi}\over {\partial\lam_i}}
+e^{2\lam_k}{{\partial^2\Phi}\over {\partial\phi_\a^2}}
+e^{2(\lam_k+\lam_l)}
  {{\partial^2\Phi}\over {\partial\phi_\b^2}}
+e^{2\lam_l}{{\partial^2\Phi}\over {\partial\phi_l^2}}
\nn\\ &&
-e^{2\lam_l}\phi_\a{{\partial^2\Phi}\over {\partial\phi_l\partial\phi_\b}}
+e^{2\lam_k}\phi_l {{\partial^2\Phi}\over {\partial\phi_\a\partial\phi_\b}}
+{1\over 4}\left(e^{2\lam_l}\phi_\a^2 + e^{2\lam_k}\phi_l^2\right)
  {{\partial^2\Phi}\over {\partial\phi_\b^2}}
\nn
\eear
Let us look for a solution that behaves like:
$$
\Phi = e^{-T(\lam,\phi_l) + 2\pi i n \phi_\a 
  + 2\pi i m (\phi_\b +{1\over 2}\phi_\a\phi_l)}.
$$
We find the equation:
\bear
0 &=&
-{{\partial^2 T}\over {\partial\lam_k^2}}
+{{\partial^2 T}\over {\partial\lam_k\partial\lam_l}}
-{{\partial^2 T}\over {\partial\lam_l^2}}
\nn\\ &&
+\left({{\partial T}\over {\partial\lam_k}}\right)^2
+\left({{\partial T}\over {\partial\lam_l}}\right)^2
-\left({{\partial T}\over {\partial\lam_k}}\right)
    \left({{\partial T}\over {\partial\lam_l}}\right)
+{{\partial T}\over {\partial\lam_k}}
+{{\partial T}\over {\partial\lam_l}}
\nn\\ &&
-4\pi^2 (n+m\phi_l)^2 e^{2\lam_k}
-e^{2\lam_l} {{\partial^2 T}\over {\partial\phi_l^2}}
+e^{2\lam_l}\left({{\partial T}\over {\partial\phi_l}}\right)^2
\nn
\eear
One solution is, as expected from U-duality:
$$
T = 2\pi  e^{2\lam_k}\sqrt{m^2 e^{2\lam_l} + (n+m\phi_l)^2}.
$$
Finally, we would like to recall the case of subsection (\ref{parspec}).
This case is particularly interesting because it gives us a glimpse
of the particle spectrum.

The intersection matrix is:

\begin{tabular}{c|rrrrr|}
         & $\a_0$ &  $\b$  &  $\g$  & $\delta$ & $\eta$ \\ \hline
$\a_0$   &    2   &    1   &    0   &    1     &    0   \\
$\b$     &    1   &    2   &   -1   &    1     &    0   \\
$\g$     &    0   &   -1   &    2   &   -1     &    0   \\
$\delta$ &    1   &    1   &   -1   &    2     &    1   \\
$\eta$   &    0   &    0   &    0   &    1     &    2   \\ \hline
\end{tabular}

It would be interesting to check that the behavior suggested
in (\ref{TzkT}) is related to a harmonic function.
We will check this in another work \cite{WIP}, but we will make a few
comments.
For the check, it seems imperative to find a realization of
$\a_0\dots\eta$ such that the set of roots that are smaller than
at least one of $\a_0\dots\eta$ has the smallest number of elements
as possible. We can then search for a harmonic function
$\Phi$ that depends only on $\phi_{\a_0},\dots,\phi_\eta$
and the phases that correspond to these extra roots,
because the extra roots will enter (\ref{Wuvxi}-\ref{Wuxi}).
We should also make sure that relations among roots such as
$\a_0 > \b$ should be preserved.

Let us ignore the extra root $\eta$ and check the simplest
version of (\ref{TzkT}) with $l=0$.
If we choose the simple roots of $E_8$ to be $\rho_1,\dots,\rho_8$
with
$$
T_{\rho_1} = R_1 R_2^{-1},
T_{\rho_2} = R_2 R_3^{-1},
\dots,
T_{\rho_7} = R_7 R_8^{-1},
T_{\rho_8} = R_6 R_7 R_8,
$$
then a minimal choice for $\a_0,\dots,\delta$ can be taken as:
\bear
\a_0   &=& \rho_5 + \rho_6 + \rho_8,\qquad T_{\a_0} = R_5 R_6 R_8,\nn\\
\b     &=& \rho_5 + \rho_8,\qquad T_{\b} = R_5 R_7 R_8,\nn\\
\g     &=& \rho_4 + \rho_5 + \rho_6,\qquad T_{\g} = R_4 R_6^{-1},\nn\\
\delta &=& \rho_8,\qquad T_{\delta} = R_6 R_7 R_8,\nn
\eear
In addition to these roots, the Laplacian depends on the following
roots:
\bear
\chi_1 &=& \rho_4,\qquad T_{\chi_1} = R_4 R_5^{-1},\nn\\
\chi_2 &=& \rho_5,\qquad T_{\chi_2} = R_5 R_6^{-1},\nn\\
\chi_3 &=& \rho_6,\qquad T_{\chi_3} = R_6 R_7^{-1},\nn\\
\chi_4 &=& \rho_4 + \rho_5,\qquad T_{\chi_4} = R_4 R_6^{-1},\nn\\
\chi_5 &=& \rho_5 + \rho_6,\qquad T_{\chi_5} = R_3 R_7^{-1},\nn
\eear
Physically, this means that the equation will depend
on five more angles ($\phi_{\chi_1},\dots,\phi_{\chi_5}$).
Note that all this roots can be embedded inside an $SO(4,4)$ subgroup.
We will not pursue this direction here, but
we note that if the conjecture at the end of
section (\ref{roots}) is correct, then since 
this instanton can be embedded as an instanton in M-theory
on $\MT{8}$ its action has to be harmonic and the expectation 
(\ref{TzkT}) would be met.


\section{Discussion}\label{disc}

There are two established facts that seem fascinating and were part
of the motivation for the conjectures presented above.
The first fact is that the actions of wrapped branes are encoded
in {\em exact} harmonic functions on
$E_{d(d)}(\BZ)\backslash E_{d(d)}(\BR)/\Kom_d$.
Precisely which brane we are asking about is encoded in the dependence
of the function on the periodic variables (the ``phases'') in the
moduli space.
It is not only the action of single wrapped BPS branes that is the
exponent of a harmonic function but, as we have seen in 
section (\ref{roots}), combinations of several
branes are also encoded in harmonic functions.
This leads one to suspect that Laplace's equation on the moduli space
is analogous to a a second-quantized equation of motion rather than
just a first-quantized equation.
If harmonic functions with a given behavior as a function of phases
encode the action of the BPS instantons, then the natural question
is what would harmonic functions with more complicated behavior,
as a function of phases, encode.

The second fascinating fact is that one can realize mass-like
deformations of gauge theories (such as giving mass to the adjoint 
hypermultiplet in $\SUSY{4}$ SYM) by studying the behavior of
the corresponding branes in the presence of another, ``spectator''
brane. If this construction is embedded inside $\MT{d}$ then
each type of brane is coupled to its own periodic phase in the moduli
space and the mass parameter corresponds to a different periodic phase.

We might also be able to realize certain nonsupersymmetric
deformations of gauge theories in this manner but one is forced
to work with M-theory on tori $\MT{d}$ with $d\ge 9$.
we can now define a harmonic function on the 
generalization of the ``moduli-space'',
$E_{d(d)}(\BZ)\backslash E_{d(d)}/\Kom_d$, and we can extract the
piece of it that has the desired behavior as a function of the phases.
The question stands: {\em what would this mode of the harmonic function
describe?}
The natural conjecture is that it will correspond to the partition 
function of the gauge theory (multiplied by the contribution
of the tensions of the branes to the action).


If that is true then it follows that there is a {\em single}
harmonic function $\ZM$ on $E_{d}(\BR)\backslash E_{d}(\BR)/\Kom$
that encodes all of the separate partition functions discussed above.
The different partition functions can be obtained from $\ZM$
by extracting particular Fourier modes of the function with respect
to appropriate periodic phases.
Here, $d$ should be large enough to accommodate the nonsupersymmetric
constructions and should probably be $d=10$, or perhaps higher!


In \cite{GanXi} we also conjectured that certain types of ``gravity''
partition functions can be defined and that they are also encoded in
$\ZM$. The conjectured partition functions were defined to be the
partition function of M-theory on a space that is constructed
as an $\MR{4}$ fibration over $\MT{7}$ and as $\MR{2}$ fibrations
over $\MT{9}$. In \cite{GanXi} we argued
that the former is a special case of the latter and we presented
a conjectured prescription for extracting the latter out of $\ZM$.
In section (\ref{higherd}) we suggested a different prescription
for extracting out of $\ZM$ the partition function of M-theory
on an $\MR{4}$ fibration over $\MT{7}$. 

I do not see how this agrees
with the prescription in \cite{GanXi} (in the special case of 
an $\MR{4}$ fibration over $\MT{7}$). The present prescription
treats both $SU(2)_L$ and $SU(2)_R$ factors of the fibration group
$SO(4)$ symmetrically. Perhaps the conjectured
prescription of \cite{GanXi} for an $\MR{2}$ fibration over $\MT{9}$
is wrong, or perhaps it somehow reduces to the present one in a nontrivial
fashion (or perhaps both are wrong!).

In this paper we showed how certain deformations of the D-instanton
actions can be realized using the phases of 
$E_{d(d)}(\BZ)\backslash E_{d(d)}/\Kom_d$. We were only able to realize
certain mass deformations and we haven't discussed deformations that
are not quadratic in the fields
(with the exception of section (\ref{mintwo})).

It would be interesting to understand what other deformations
of the D-instanton integral are possible using hooks and baits.
In \cite{Myers}, the effect of RR fields on the D0-brane
action was studied. It was found that certain fields induce
quartic and higher terms.
It would also be interesting to understand the behavior of the D0-brane
or D-instanton
in nonzero RR field strengths. Such configurations can arise
when the D-instanton ``probes'' other objects.
In a related paper \cite{TayRam}, the coupling of closed string states
to the D0-brane action was studied and many
augmentations of the D0-brane action can arise this way.
Although the closed string fields
are not directly related to moduli and therefore variables of
$E_8$, perhaps one can turn them on by adding more hooks and baits.
One would probably have to utilize the
other roots of $E_{10}$ (perhaps even
the ``imaginary'' roots that satisfy $\lrp{\a}{\a}\le 0$).

We have argued in section (\ref{groups}) that
in many cases one can separate from the infinite dimensional $E_{10}$
a subset of a finite number of variables that are relevant for
the problem and ignore the rest.
We have suggested that the conjecture can therefore be tested in
certain cases for which we know the existence of BPS states in 
the spectrum. Thus, one can test the conjecture that
(\ref{TzkT}) is a limit of a harmonic function.
Similarly, the mass formulas in sections (\ref{twhook})
might be tested along similar lines.

The fact that
one can separate from the infinite dimensional $E_{10}$
a subset of a finite number of variables
seems suspicious at first, because we do not expect a 
differential equation in a finite number of variables to encode
the partition functions of complicated gauge theories.
However, first of all, I do not see an immediate contradiction.
Moreover, given any partition function $Z(R_1,R_2,R_3,R_4)$,
one can always add even just one single variable
say $\lam$ to the existing list of variables and find
a harmonic function $Z(R_1,R_2,R_3,R_4,\lam)$ that reduces to
$Z(R_1,R_2,R_3,R_4)$ in the limit $\lam\rightarrow\infty$.
In the context of $E_{10}$, it would seem that the heart of the matter
is the boundary conditions imposed on $\Xi$ by the U-duality group
$E_{10}(\BZ)$.

We have suggested that, if one adds more variables, the partition function
of the D-instanton matrix integrals can be read off from a harmonic equation.
It is actually a well known fact that if one adds all the deformations
to the matrix integral one can write down a set of second-order differential
equations (see \cite{AvJe} for the case with many matrices).
We start with:
$$
Z(\{\sigma_{\mu_1,\dots,\mu_r}\})
=\sum_N\int \prod_{\u=1}^{10} dX_\u {\mbox{exp}}\,{\mbox{Tr}}\left\{
-\sum \sigma_{\mu_1,\dots,\mu_r}X_{\mu_1}\cdots X_{\mu_r}
\right\}.
$$
In particular, $\sigma$ (with no indices) is the coefficient of 
$N={\mbox{Tr}}1$.
To get a second-order equation we insert:
\be
\sum_N\int \prod_{\u=1}^{10} dX_\u\,
{\mbox{Tr}}\left\lbrack
X_{\v_1}\cdots X_{\v_s}\ppx{X_\u}
\right\rbrack
{\mbox{exp}}\,{\mbox{Tr}}\left\{
-\sum \sigma_{\u_1,\dots,\u_r}X_{\u_1}\cdots X_{\u_r}
\right\}
\ee
and integrate by parts.
We find:
\be
\left\lbrack
\sum_j \sigma_{\u_1,\dots,\u_r\u}
  \ppx{\sigma_{\v_1,\cdots \v_s \u \u_1,\dots,\u_{r}}}
-\sum_j\delta_{\u\v_j}
 \ppypxpx{}{\sigma_{\v_1,\cdots \v_{j-1}}}{\sigma_{\v_{j+1},\cdots\v_s}}
\right\rbrack Z = 0.
\ee
This is an infinite set of equations.
Perhaps they are somehow related to the harmonic equation on $\ZM$
by expanding in certain regions of moduli space
where a certain subset of variables decouple. Expanding the equation 
in these variables might lead to an infinite set of second order
equations.


One aspect of the conjecture about $\ZM$ is that the U-duality
$E_{10}(\BZ)$ plays an important role.
It is not hard to find harmonic functions that reduce to any function
that we want once we take a limit of a certain variable.
After all, we can solve Laplace's equation with any given initial 
condition. What is nontrivial, is to find harmonic functions
which satisfy certain boundary conditions.
In our case the boundary conditions are set by $E_{10}(\BZ)$ that
relate various regions of the 10 noncompact parameters  of $E_{10}$.
It is reasonable to expect that $\ZM$ should vanish in
the limits that correspond to decompactification of enough
(probably 3 or more) directions and all their U-dual limits.
However, not all limits of the $E_{10}$ moduli space
can be obtained this way \cite{BFM}. It is not clear to me
what the behavior of $\ZM$ should be in those other regions.


Another, perhaps interesting, direction for research would be to
further explore the combination of instantons that correspond
to roots $\a$ and $\b$ with $\lrp{\b}{\a} = -2$.
This seems to be the next interesting case (in the sense that
supersymmetry is preserved) after the case of $\lrp{\b}{\a}=0$.
Among the examples presented in section (\ref{mintwo})
are various exotic combinations of KK monopoles and branes
that may lead to interesting physical effects.

There are also interesting systems that are U-dual to the well studied
cases. One is the behavior of D-branes, M2-branes and M5-branes
near the core of KK-monopoles with NSNS or 3-form fluxes turned on.
This was discussed briefly in subsection (\ref{InstKK}) and
will hopefully be explored further somewhere else \cite{WIP}.
Another system that will be explored further somewhere else
is the system of surfaces in $\MT{8}$ and its U-dual manifestations
that was discussed briefly in (\ref{surfT}).


We have seen that the usual questions about partition functions of
gauge theories and gravity are only a part of the answers 
that $\ZM$ is conjectured to encode. $\ZM$ probably 
contains other information about questions
that we cannot formulate in terms of space and time.


\section*{Acknowledgments}
I have benefited from various discussions with S. Sethi and W. Taylor
on issues related to this paper.


\def\np#1#2#3{{\it Nucl.\ Phys.} {\bf B#1} (#2) #3}
\def\pl#1#2#3{{\it Phys.\ Lett.} {\bf B#1} (#2) #3}
\def\physrev#1#2#3{{\it Phys.\ Rev.\ Lett.} {\bf #1} (#2) #3}
\def\prd#1#2#3{{\it Phys.\ Rev.} {\bf D#1} (#2) #3}
\def\ap#1#2#3{{\it Ann.\ Phys.} {\bf #1} (#2) #3}
\def\ppt#1#2#3{{\it Phys.\ Rep.} {\bf #1} (#2) #3}
\def\rmp#1#2#3{{\it Rev.\ Mod.\ Phys.} {\bf #1} (#2) #3}
\def\cmp#1#2#3{{\it Comm.\ Math.\ Phys.} {\bf #1} (#2) #3}
\def\mpla#1#2#3{{\it Mod.\ Phys.\ Lett.} {\bf #1} (#2) #3}
\def\jhep#1#2#3{{\it JHEP.} {\bf #1} (#2) #3}
\def\atmp#1#2#3{{\it Adv.\ Theor.\ Math.\ Phys.} {\bf #1} (#2) #3}
\def\jgp#1#2#3{{\it J.\ Geom.\ Phys.} {\bf #1} (#2) #3}
\def\cqg#1#2#3{{\it Class.\ Quant.\ Grav.} {\bf #1} (#2) #3}
\def\hepth#1{{\it hep-th/{#1}}}



\begin{thebibliography}{99}

\bibitem{GanXi}{O.J. Ganor,
  {``Two Conjectures on Gauge-Theories, Gravity,
  and Infinite Dimensional Kac-Moody Groups,''}
  PUPT-1836, \hepth{9903110}}

\bibitem{BBS}{K. Becker, M. Becker and A. Strominger,
  {``Five-Branes, Membranes and Non-Perturbative String Theory''} 
  \np{456}{1995}{130-152}, \hepth{9507158}}

\bibitem{WitNON}{E. Witten,
  {``Nonperturbative Superpotentials In String Theory''},
  \np{474}{1996}{343-360}, \hepth{9604030}}

\bibitem{WitBND}{E.~Witten,
  {``Bound States of Strings and $p$-Branes,''}
  \np{460}{1996}{335--350}, \hepth{9510135}}

\bibitem{StrOPN}{A. Strominger,
  {``Open p-Branes,''} \hepth{9512059}}

\bibitem{SeiVBR}{N.~Seiberg,
  {``New Theories in Six-Dimensions and
  Matrix Description of M-theory on $T^5$ and $T^5/Z_2$,''}
  \hepth{9705221}, \pl{408}{1997}{98}}


\bibitem{BDS}{T. Banks, M.R. Douglas and N. Seiberg,
  {``Probing F-theory With Branes,''}
  \pl{387}{1996}{278--281}, \hepth{9605199}}

\bibitem{SeiIRD}{N. Seiberg,
  {``IR Dynamics on Branes and Space-Time Geometry,''}
  \pl{384}{1996}{81--85}, \hepth{9606017}}

\bibitem{CGKM}{Y.-K.E.~Cheung, O.J.~Ganor,
  M.~Krogh and A.Yu.~Mikhailov, {``Noncommutative Instantons and
 Twisted (2,0) and Little String Theories,''} \hepth{9812172}}

\bibitem{WitBR}{E.~Witten,
  {``Solutions Of Four-Dimensional Field Theories Via M Theory,''}
  \np{500}{1997}{3--42},\hepth{9703166}}

\bibitem{Douglas}{M.R.~Douglas,
  {``Branes within Branes,''} \hepth{9512077}}

\bibitem{BSSil}{T.~Banks, N.~Seiberg and E.~Silverstein,
  {``Zero and One-dimensional Probes with N=8 Supersymmetry,''}
  \pl{401}{1997}{30}, \hepth{9703052}}

\bibitem{DFK}{U.H.~Danielsson, G.~Ferretti and I.R.~Klebanov,
  {``Creation of Fundamental Strings by Crossing D-branes,''}
  \physrev{79}{1997}{1984}, \hepth{9705084}}

\bibitem{BGL}{O.~Bergman, M.R.~Gaberdiel, G.~Lifschytz,
  {``Branes, Orientifolds and the Creation of Elementary Strings,''}
  \np{509}{1998}{194}, \hepth{9705130}}

\bibitem{OPRev}{N.A. Obers and B. Pioline,
  {``U-duality and M-theory,''} \hepth{9809039}}
\bibitem{OPRevSh}{N.A. Obers and B. Pioline,
  {``U-duality and M-theory, an algebraic approach,''} \hepth{9812139}}
\bibitem{OPNew}{N.A. Obers and B. Pioline,
  {``Einstein Series and String Thresholds,''} \hepth{9903113}}

\bibitem{GGI}{M.B.~Green and M.~Gutperle,
   {``$\lambda^{16}$ and related terms in M-theory on $T^2$,''}
   \hepth{9710151}}

\bibitem{PiolH}{B. Pioline,
  {``A Note on nonperturbative $R^4$ couplings,''}
  \pl{431}{1998}{73}, \hepth{9804023}}

\bibitem{GreSet}{M.B.~Green and S.~Sethi,
  {``Supersymmetry Constraints on Type-IIB Supergravity,''}
  \hepth{9808061}}

\bibitem{BerVaf}{N.~Berkovits and C.~Vafa,
  {``Type-IIB $R^4 H^{4g-4}$ Conjectures,''}
  \np{533}{1998}{181}, \hepth{9803145}}

\bibitem{EGKR}{S.~Elitzur, A.~Giveon, D.~Kutasov, E.~Rabinovici,
  {``Algebraic Aspects of Matrix Theory on $T^d$,''}
  \hepth{9707217}}

\bibitem{WatZS}{W. Taylor IV,
  {``Adhering 0-branes to 6-branes and 8-branes,''}
  \np{508}{1997}{122}, \hepth{9705116}}

\bibitem{Kac}{V.G.~Kac,
  {\em ``Infinite dimensional Lie algebras,''}
  Cambridge University Press, 1990, 3rd edition}

\bibitem{SenUD}{A. Sen,
  {``U-duality and Intersecting D-branes,''}
  \physrev{53}{1996}{2874-2894}, \hepth{9511026}}

\bibitem{SenMA}{A. Sen,
  {``A Note on Marginally Stable
     Bound States in Type II String Theory,''}
  \physrev{54}{1996}{2964-2967}, \hepth{9510229}}

\bibitem{BerKol}{O. Bergman and B. Kol,
  {``String webs and ${1\over 4}$-BPS Monopoles,''}
  \np{536}{1998}{149-174}, \hepth{9804160}}

\bibitem{CDS}{A.~Connes, M.R.~Douglas and A.~Schwarz,
  {``Noncommutative Geometry and Matrix Theory:
  Compactification on Tori,''} \jhep{02}{1998}{003}, \hepth{9711162}}

\bibitem{DH}{M.R.~Douglas and C.~Hull,
  {``D-branes and the Noncommutative Torus,''}
  \jhep{02}{1998}{008}, \hepth{9711165}} 

\bibitem{SWncg}{N.~Seiberg and E.~Witten,
  ``String Theory and Non-Commutative Geometry,'' \hepth{9908142}}

\bibitem{WIP}{Work in progress}

\bibitem{LPTi}{J. Lykken, E. Poppitz and S.P. Trivedi,
  {``Chiral Gauge Theories From D-Branes,''}
  \pl{416}{1998}{286}, \hepth{9708134}}

\bibitem{HU}{A. Hanany and A.M. Uranga,
  {``Brane Boxes and Branes on Singularities,''}
  \hepth{9805139}}

\bibitem{GanZ}{O.J. Ganor,
  {``A Note On Zeroes of Superpotentials in F-Theory,''}
  PUPT-1672, \np{499}{1997}{55}, \hepth{9612077}}

\bibitem{PolWit}{J. Polchinski and E. Witten,
  {``Evidence For Heterotic - Type I String Duality''},
  \hepth{9510169}}

\bibitem{SenD}{A. Sen,
  {``Dynamics of Multiple Kaluza-Klein Monopoles in 
  M and String Theory,''}
  \atmp{1}{1998}{115-126}, \hepth{9707042}}

\bibitem{WitADE}{E. Witten,
  {``Heterotic String Conformal Field Theory and ADE Singularities,''}
  \hepth{9909229}}


\bibitem{KKN}{V. Kazakov, I. Kostov and N. Nekrasov,
  {``D-Particles, Matrix Integrals and KP Hierarchy,''}
  \hepth{9810035}}


\bibitem{GriHar}{Griffiths and Harris,
   {\it ``Principles of Algebraic Geometry''},
   Wiley-Interscience, New-York, 1978}

\bibitem{Myers}{R.C. Myers,
   {``Dielectric Branes,''} \hepth{9910053}}

\bibitem{TayRam}{W. Taylor and M. van Raamsdonk,
  {``Multiple D0-branes in Weakly Curved Backgrounds,''}
  \hepth{9904095}}

\bibitem{AvJe}{J.~Avan and A.~Jevicki,
    {``Collective Hamiltonians with Kac-Moody Algebraic Conditions,''}
    \np{439}{1995}{679-691}, \hepth{9410166}}

\bibitem{BFM}{T.~Banks, W.~Fischler and L.~Motl,
  {``Dualities versus Singularities,''} \hepth{9811194}}


\end{thebibliography}
\end{document}